%% file: ms.tex
\documentclass[a4paper,11pt]{article}
\pdfoutput=1 
             
\usepackage{jcappub}
\usepackage{graphicx}
\usepackage{color}
\usepackage{makecell}
\usepackage{afterpage}
\usepackage[dvipsnames]{xcolor}
\usepackage{multirow}
\usepackage{cleveref}
\usepackage[normalem]{ulem}		
\usepackage{xspace}
\usepackage{subcaption}
\usepackage{booktabs}
\usepackage{comment}
\usepackage{physics} 

\usepackage{centernot}
\usepackage{mathtools}
\usepackage{stmaryrd}
\usepackage[utf8]{inputenc}

\usepackage{placeins}
\usepackage{titlesec}
\usepackage{csquotes}
\usepackage{enumitem}
\titlespacing*{\section} {0pt}{1.5ex plus 0.5ex minus .2ex}{-1.8ex plus .2ex}
\titlespacing*{\subsection} {0pt}{0.75ex plus 0.5ex minus .2ex}{-1.8ex plus .2ex}

\input{Commands.tex}

\title{Cosmological constraints on multi-interacting dark matter}

\author[1]{Niklas Becker,}
\author[2]{Deanna C. Hooper,}
\author[1]{Felix Kahlhoefer,}
\author[1]{Julien Lesgourgues,} 
\author[1]{and Nils Sch\"oneberg}

\affiliation[1]{Institute for Theoretical Particle Physics and Cosmology (TTK), \\ RWTH Aachen University, D-52056 Aachen, Germany.}
\affiliation[2]{Service de Physique Th\'eorique, CP225, \\ Universit\'e Libre de Bruxelles, Boulevard du Triomphe, 1050 Bruxelles, Belgium.}

\emailAdd{nrbecker@physik.rwth-aachen.de}
\emailAdd{deanna.hooper@ulb.be}
\emailAdd{kahlhoefer@physik.rwth-aachen.de}
\emailAdd{lesgourg@physik.rwth-aachen.de}
\emailAdd{schoeneberg@physik.rwth-aachen.de}

\abstract{
The increasingly significant tensions within $\Lambda$CDM, combined with the lack of detection of dark matter (DM) in laboratory experiments, have boosted interest in non-minimal dark sectors, which are theoretically well-motivated and inspire new search strategies for DM. Here we consider, for the first time, the possibility of DM having simultaneous interactions with photons, baryons, and dark radiation (DR). We have developed a new and efficient version of the Boltzmann code \class that allows for one DM species to have multiple interaction channels. With this framework we reassess existing cosmological bounds on the various interaction coefficients in multi-interacting DM scenarios. We find no clear degeneracies between these different interactions and show that their cosmological effects are largely additive. We further investigate the possibility of these models to alleviate the cosmological tensions, and find that the combination of DM--photon and DM--DR interactions can at the same time reduce the $S_8$ tension (from $2.3\sigma$ to $1.2\sigma$) and the $H_0$ tension (from $4.3\sigma$ to $3.1\sigma$). The public release of our code will pave the way for the study of various rich dark sectors.
}

\begin{document}

\hfill{\small TTK-20-32}

\hfill{\small ULB-TH/20-13}

\vspace{-2\baselineskip}

\maketitle

\setlength{\parskip}{\baselineskip}%
\input{Introduction.tex}

\input{Theory.tex}
\input{Results.tex}
\input{Tension.tex}

\input{Discussion.tex}

\newpage
\section*{Acknowledgements}
We thank Torsten Bringmann for very helpful discussions, Sebastian Bohr for comments on the DM--DR interactions, and Vera Gluscevic for feedback on the draft. DH is supported by the FNRS research grant number \mbox{F.4520.19}. FK is supported by the DFG Emmy Noether Grant No.\ KA 4662/1-1. JL is supported by the DFG grant LE 3742/3-1. NS acknowledges support from the DFG grant LE 3742/4-1. Simulations were performed with computing resources granted by RWTH Aachen University under project jara0184 and thes0811.

\appendix
\input{App_Equation.tex}
\input{App_Decoupling.tex}
\input{App_Implementation.tex}

\bibliography{biblio}{}
\bibliographystyle{JHEP}

\end{document}

%% file: Commands.tex
\newcommand{\stkout}[1]{\ifmmode\text{\sout{\ensuremath{#1}}}\else\sout{#1}\fi}

\definecolor{amber}{rgb}{1.0, 0.75, 0.0}

\newcolumntype{L}[1]{>{\raggedright\arraybackslash}p{#1}}
\newcolumntype{C}[1]{>{\centering\arraybackslash}p{#1}}

\newcommand{\class}{{\sc class}\xspace}

\newcommand{\lcdm}{$\Lambda$CDM}

\newcommand{\DM}{\mathrm{DM}}
\newcommand{\DR}{\mathrm{DR}}
\newcommand{\g}{\gamma}
\renewcommand{\H}{\mathcal{H}}

%% file: Introduction.tex
\section{Introduction}\label{sec:intro}
The cold dark matter (CDM) paradigm, which assumes cold and collisionless dark matter (DM) particles interacting only gravitationally, is a cornerstone of both cosmology and particle physics. This scenario is supported by a wide range cosmological observations at many different epochs, including CMB missions~\cite{Akrami:2018vks}, BAO data~\cite{Beutler:2011hx,Ross:2014qpa,Alam:2016hwk}, observations of galaxy clusters~\cite{Clowe:2006eq}, and weak lensing experiments~\cite{Heymans:2013fya,Abbott:2020knk,Joudaki:2019pmv}.

Despite the overwhelming success of CDM, and by extension of the standard \lcdm~cosmological model, in recent years possible tensions have become more apparent. The most well-known of these is the Hubble tension, whereby the expansion rate of the universe (quantified with $H_0$) as inferred by CMB~\cite{Akrami:2018vks} and BAO~\cite{Beutler:2011hx,Ross:2014qpa,Alam:2016hwk} measurements differs by more than $4.4\sigma$  from that measured in the local universe~\cite{Knox:2019rjx,Riess:2019cxk,Verde:2019ivm} 

Moreover, the clustering of matter on scales of $\sim8 \,\mathrm{Mpc/h}$ (quantified with $S_8$) inferred from CMB data~\cite{Akrami:2018vks} is in more than $2\sigma$ tension~\cite{MacCrann:2014wfa, Chang:2018rxd} with the measurements obtained from weak lensing experiments~\cite{Heymans:2013fya,Asgari:2019fkq,Abbott:2020knk,Joudaki:2019pmv,2020arXiv200715632H} -- this is known as the $S_8$ tension. Furthermore, there are possible shortcomings of CDM when looking at structure formation on small scales~\cite{Flores:1994gz,Moore:1994yx,Klypin:1999uc,deBlok:2009sp,BoylanKolchin:2011de,BoylanKolchin:2011dk, Oman:2015xda,Kamada:2016euw,Tulin:2017ara,Salucci:2018hqu}  
Finally, the observation by the EDGES collaboration of a colder 21cm spin temperature than expected~\cite{Bowman:2018yin} further called into question the CDM paradigm~\cite{Barkana:2018lgd} (see however Refs.~\cite{Munoz:2018pzp,Berlin:2018sjs,Barkana:2018qrx} for further discussions on this interpretation). 

These issues, combined with the lack of detections in DM experiments, have motivated interest in models beyond the standard CDM paradigm, such as Interacting Dark Matter (IDM). These interacting scenarios can be broadly separated into two categories: interactions within the dark sector, such as Self Interacting Dark Matter~\cite{Spergel:1999mh,ArkaniHamed:2008qn,Feng:2009mn,Feng:2009hw,Buckley:2009in,Tulin:2017ara} or DM interacting with an additional relativistic species 
(Dark Radiation, henceforth DR)~\cite{Archidiacono:2011gq, Diamanti:2012tg, Cyr-Racine:2013fsa,Chu:2014lja,Rossi:2014nea,Buen-Abad:2015ova,Lesgourgues:2015wza,Cyr-Racine:2015ihg,Schewtschenko:2015rno,Krall:2017xcw,Archidiacono:2017slj,Buen-Abad:2017gxg,Archidiacono:2019wdp}; and interactions between DM and Standard Model particles like 
baryons~\cite{Chen:2002yh,Boehm:2004th,Sigurdson:2004zp,Melchiorri:2007sq,ArkaniHamed:2008qn,Dvorkin:2013cea,Ali-Haimoud:2015pwa,Munoz:2015bca,Kadota:2016tqq,Munoz:2017qpy,Gluscevic:2017ywp,Boddy:2018kfv,Ali-Haimoud:2018dvo,Barkana:2018lgd,Boddy:2018wzy,Bringmann:2018cvk,Emken:2018run,Slatyer:2018aqg,Xu:2018efh}, 
photons~\cite{Boehm:2004th,Weiner:2012cb,Wilkinson:2013kia,Boehm:2014vja,Ali-Haimoud:2015pwa,Diacoumis:2018ezi,Escudero:2018thh,Kumar:2018yhh,Stadler:2018jin}, 
or neutrinos~\cite{Bringmann:2013vra,Audren:2014lsa,Cherry:2014xra,Wilkinson:2014ksa,Horiuchi:2015qri,Ghosh:2017jdy,Diacoumis:2018ezi,Campo:2018dfh,DiValentino2018a,Olivares-DelCampo2018,Pandey:2018wvh,Choi:2019ixb, Stadler:2019dii}. 

While these individual interactions have been studied extensively in the literature (see references above), in this paper we aim, for the first time, to study scenarios in which the IDM has several interactions simultaneously. 
Such multiple interactions are generically expected to be present if the DM particle is part of a larger dark sector with several new states. For example, if DM interacts with DR in the form of sterile neutrinos, the same mediator that induces DM--DR interactions may also generate DM--baryon interactions. Alternatively, if DM interacts with DR in the form of massless dark photons, mixing between the dark and visible photon may give rise also to DM--photon interactions, which then in turn induce DM--baryon interactions.
The goal of this paper is three-fold: first we will develop the formalism needed in order to describe these simultaneous interactions, which require non-trivial modifications,  such as  for the temperature evolution of the different species. 
Second, we will assess the cosmological bounds on different IDM cross sections in models with two or three simultaneous interactions. We wish to check whether they differ from those obtained with single interactions,
since in principle, some cancellations between the various effects could lead to parameter degeneracies. 
Third, we will study the possible implications these multi-interacting scenarios have on the aforementioned cosmological tensions. 

In order to do this, we have developed a new version of the Boltzmann solver \class~\cite{Blas:2011rf} featuring DM-DR interactions (already present since \class v2.9~\cite{Archidiacono:2017slj, Archidiacono:2019wdp}), DM-baryon interactions, and DM-photon interactions in a unified and systematic approach, without substantial increase of the runtime. The code developed here will be made publicly available in a forthcoming release, \class v3.1. 

This paper is organised as follows. In Sec.~\ref{sec:theory} we review the different interacting models we will consider in this work, highlighting in Secs.~\ref{sec:th_idm}-\ref{sec:th_tca} the important considerations needed when combining these into multi-interacting models. In Sec.~\ref{sec:th_imp} we illustrate the effects of these interactions on the cosmological observables. In Sec.~\ref{sec:res} we present our results first for the single interaction models (Sec.~\ref{sec:res_single}), and then for all possible dual or triple interaction scenarios (Sec.~\ref{sec:res_dual}).
Additionally, we provide a detailed description of all the relevant equations in App.~\ref{App:eqs}, a calculation of the relevant decoupling redshifts in App.~\ref{App:dec}, and details on the numerical implementation in \class in App.~\ref{App:num}.

%% file: Theory.tex
\section{Dark matter interactions}
\label{sec:theory}
\enlargethispage{2\baselineskip}
In this section we review the different kinds of DM interactions considered in this work. We emphasize that we describe these interactions at an effective level in the form of temperature-dependent cross sections, rather than at a fundamental level in the form of Lagrangian densities. This means, in particular, that we will treat different types of interactions as independent, even though they may be linked in a fundamental theory. Nevertheless, we will take inspiration from particle physics to identify particularly well-motivated scenarios and comment on potential complementary constraints below.

We highlight that in this work and in the forthcoming \class v3.1 release, we only consider one single IDM species with potentially all of the relevant interaction channels. From a numerical point of view, this case is the easiest one to generalise if one wishes to study interacting dark sector models with an arbitrary degree of complexity. Indeed, one could very easily nest the new lines of code relative to IDM species inside loops over $N_\mathrm{IDM}$ different species, as done in \class for non-cold dark matter. This would result in an arbitrary number of IDM species whose interaction channels could be switched on and off independently. For instance, one could have part of DM interacting with baryons, and part of it interacting with dark radiation. In principle, such a model could slightly differ from the case of a single DM particle interacting with both species that we consider in this work.

\subsection{Dark matter -- baryon interactions}
\label{sec:th_dmb} 

Scattering between DM and baryons can lead to an exchange of momentum proportional to the momentum transfer cross section
\begin{equation}
 \sigma_\mathrm{T} = \int \mathrm{d}\Omega \frac{\mathrm{d}\sigma}{\mathrm{d}\Omega} (1 - \cos \theta) \, .
\end{equation}
In weakly-coupled theories, $\sigma_\mathrm{T}$ can only depend on even powers of the DM--baryon relative velocity $v$ and in many cases this dependence is given by a power law.\footnote{It has been argued that non-perturbative effects corresponding to the temporary formation of DM--baryon bound states could give rise to additional factors depending on odd powers of the relative velocity. However, to the best of our knowledge, no such model has been worked out in detail and hence we do not include this possibility in the present work. We also do not consider the possibility that $\sigma_\mathrm{T}$ depends logarithmically on $v$, which can occur in models with long-range interactions.} In the present work we consider $\sigma_\mathrm{T} = \sigma_{\DM\text{--}b} v^{n_b}$  with $n_b = \left\lbrace -4, -2, 0\right\rbrace$. The case $n_b=-4$ arises, for example, in models of DM with a fractional electric charge~\cite{Melchiorri:2007sq}, which have received much interest recently in attempts to explain the EDGES anomaly~\cite{Bowman:2018yin} (see e.g.\ Ref.~\cite{Berlin:2018sjs,Kovetz:2018zan}), while the cases $n_b = -2$ and $n_b = 0$ occur in models with DM dipole moments~\cite{Sigurdson:2004zp} and contact interactions~\cite{Chen:2002yh}, respectively. Positive powers of $v$ are also possible, but lead to an interaction that becomes irrelevant at low temperatures and hence is of limited interest for this study.

We have implemented DM--baryon interactions in \class following the formalism described in Refs.~\cite{Dvorkin:2013cea, Munoz:2015bca, Xu:2018efh, Slatyer:2018aqg}, among others. Within this framework, it is assumed that both DM and baryons are non-relativistic (valid for DM masses above the MeV scale), and that in the early universe both species follow a Maxwell velocity distribution (although recently a new formalism was derived in Ref.~\cite{Ali-Haimoud:2018dvo} extending this to a general distribution via the use of the Fokker-Planck formalism). With these assumptions, the DM Euler equation (shown in full in App.~\ref{App:eqs}) will gain an additional term
\begin{equation}
\theta_\DM^\prime = \theta_\mathrm{DM, standard}^\prime - \Gamma_{\DM\text{--}b} \left( \theta_\DM - \theta_b \right) \,,
\label{eq:addboltzdmb}
\end{equation}
where $\Gamma_{\DM\text{--}b}$ is the \emph{conformal DM--baryon momentum exchange rate}, which will also appear in the modified baryon Boltzmann equations (and is called $R_\chi$ in e.g., Ref.~\cite{Dvorkin:2013cea}). Throughout this work, primes stand for derivatives with respect to conformal time. Conformal rates are defined with respect to conformal time. To quickly assess whether a given rate is efficient on cosmological time scales, one should compare it to the conformal Hubble rate $\H=a'/a$, related to the usual Hubble rate by $\H=aH$.
The rate $\Gamma_{\DM\text{--}b}$ is given by the deceleration of the DM bulk velocity. At leading order in the non-relativistic expansion, it reads
\begin{equation}
\Gamma_{\DM\text{--}b} =  \frac{a \rho_b \sigma_{\DM\text{--}b}  c_{n_b} }{m_\DM + m_b} \left( \frac{T_b}{m_b} + \frac{T_\DM}{m_\DM} + \frac{V^2_\mathrm{RMS}}{3}\right)^{\frac{n_b+1}{2}} \mathcal{F}_{He} \, ,
\label{eq:coldmb}
\end{equation}
where $T_x$ and $m_x$ represent the temperature and mass of species $x$, and $\sigma_{\DM\text{--}b}$ is the DM--baryon cross section. 
In this work we focus on scattering only with hydrogen atoms, as this is the most conservative choice. This requires setting the corrective factor $\mathcal{F}_\mathrm{He}$ to $1-Y_p \approx 0.76$ \cite{Xu:2018efh} and the average baryon mass $m_b$ to be equal to the proton mass $m_p \approx 0.938\mathrm{GeV}/c^2$\,. Our approach can be generalised to include Helium scattering as in Ref.~\cite{Dvorkin:2013cea}, while electron scattering is discussed below. 
The integration constant $c_{n_b}$  depends only on $n_b$ and is given by equation~(10) of Ref.~\cite{Dvorkin:2013cea} (for the cases most studied here, $c_{-4}=0.27$, $c_{-2}=0.53$, $c_0=2.1$).

The velocity term appearing in equation~(\ref{eq:coldmb}) is the averaged value of the DM bulk velocity relative to the baryon fluid, which is not negligible when compared to the thermal velocities of the two interacting species, thus leading to a non-linear dependence of the drag force on the DM--baryon relative velocity, as the linear theory breaks down for redshifts smaller than $z\sim10^4$. Within the formalism presented here, an approximation is made to extend the validity of the linear theory to lower redshifts. The final bulk velocity dispersion is then given by
\begin{equation}
V^2_\mathrm{RMS} \equiv \langle V^2_\DM \rangle 
\simeq  \begin{cases} 10^{-8}, & z>10^3 \\ 10^{-8} \left(\frac{(1+z)}{10^3}\right)^2,  & z\leq 10^3 \end{cases} \, .
\end{equation}
Note that an improved treatment of this relative bulk velocity was recently proposed in Ref.~\cite{Boddy:2018wzy}, which we have not included here, but which we will incorporate in a future version of our code.

For typical models, at high redshift the parenthesis in equation~(\ref{eq:coldmb}) is dominated either by ${T_b}/{m_b}$ (if DM is decoupled and $T_\DM$ remains tiny) or by the sum ${T_b}/{m_b}+T_\DM/m_\DM$ (if DM is strongly coupled with $T_\DM \simeq T_b$). In both cases, this term scales initially like $T_b \simeq T_\gamma \propto (1+z)$. It is then easy to see that $\Gamma_{\DM\text{--}b}$ scales like $(1+z)^\frac{n_b+5}{2}$, while during radiation domination the conformal Hubble rate $\H$ scales like $(1+z)$. Thus, for $n_b>-3$, DM--baryon interactions are more important at early times, and for $n_b<-3$ at late times.
The limiting case $n_b=-3$ has no special physical motivation, but it is interesting to note that it would correspond to a constant momentum exchange efficiency, since one would have $\Gamma_{\DM\text{--}b} \propto \H \propto (1+z)$ during radiation domination. 

For $n_b>-3$, we can estimate the time of DM decoupling from baryons by equating the expression of $\Gamma_{\DM\text{--}b}$ (in the limit $T_\DM \simeq T_b \simeq T_\gamma$) and $\H$. A more detailed calculation is provided in App.~\ref{App:dec}. For $n_b=-2$ the decoupling redshift is given by
\begin{equation}
	1+z_{\DM\text{--}b}^{n_b=-2} = 1.19\times10^7 \,
	\frac{(1+R_\DM)^3}{R_\DM}
	\left(
	\frac{1+ N_\mathrm{eff}f_{1\nu}}{1+3.044\, f_{1\nu}}
	\right)\!
	\left(\frac{\omega_b}{0.0224}\cdot
	\frac{\mathcal{F}_\mathrm{He}}{0.76}\cdot
	\frac{\sigma_{\DM\text{--}b}}{10^{-33} \text{cm}^2}\right)^{-2} , 
	\label{eq:zdecb2}
\end{equation}
while for $n_b=0$ we find
\begin{equation}
	1+z_{\DM\text{--}b}^{n_b=0} = 1.07\times10^5 \,
	R_\DM^{1/3} (1+R_\DM)^{1/3}
	\left(
	\frac{1+ N_\mathrm{eff}f_{1\nu}}{1+3.044\, f_{1\nu}}
	\right)^\frac{1}{3}\!
	\left(\frac{\omega_b}{0.0224} \cdot
	\frac{\mathcal{F}_\mathrm{He}}{0.76} \cdot
	\frac{\sigma_{\DM\text{--}b}}{10^{-25} \text{cm}^2}\right)^{-\frac{2}{3}}  ,
	\label{eq:zdecb0}
\end{equation} 
where we have introduced the mass ratio $R_\DM= m_\DM/m_b$ and the neutrino-to-photon density ratio (in the instantaneous decoupling limit) $f_{1\nu} = \frac{\rho_{1\nu}}{\rho_{\gamma}} =  \frac{7}{8}\left(\frac{4}{11}\right)^{4/3} \approx 0.23$.
Note that we always use a reference cross section of $\sigma_{\DM\text{--}b} \sim 10^{4n_b-25}$cm$^2$, which is the order of magnitude of the CMB bounds found in the result section (Sec.~\ref{sec:res}) for all considered values of $n_b$\,.
These results show that for allowed models with $n_b>-3$, DM always decouples from baryons during radiation domination. Since the baryon--DM momentum exchange rate is given by
\begin{equation}
\Gamma_{b\text{--}\DM} = \frac{\rho_\DM}{\rho_b} \Gamma_{\DM\text{--}b} \, ,
\end{equation}
with a ratio ${\rho_\DM}/{\rho_b}$ of order one (as long as interacting DM accounts for all or at least a sizeable fraction of DM), the same conclusions apply to the time of baryons decoupling from DM.
For $n_b=-4$,  an estimate of the ratio ${\Gamma_{\DM\text{--}b}}/{\H}$ at $z\sim10^4$ (see equation~(\ref{eq:decb4})) shows that for any cross section allowed by typical CMB bounds, $\sigma_{\DM\text{--}b} \leq {\cal O}(10^{-41}) \, \mathrm{cm}^2$, DM may recouple to baryons at the earliest around the time of photon-baryon decoupling, when $z\sim{\cal O}(10^3)$.

An important feature of these interactions is that they will substantially modify the baryon and DM temperature evolution, such that $T_\text{DM}$ needs to be numerically evolved alongside $T_b$ and $x_e$.
Since for $n_b>-3$, DM--baryon interactions couple the baryon and DM temperatures efficiently at early times, we take $T_\DM = T_b$ as the initial condition. For $n_b<-3$, on the other hand, the interactions are negligible at early times, allowing us to assume, like previous authors (see e.g. \cite{Slatyer:2018aqg}), an initial temperature $T_\DM \simeq 0$. 

We emphasize that there are, of course, strong complementary constraints on DM--baryon interactions from laboratory experiments. 
For the case of contact interactions ($n_b = 0$) the cross sections required to give interesting cosmological signals are many orders of magnitude larger than those probed by underground direct detection experiments. However, these cross sections are actually so large that the DM particles would be unable to penetrate the Earth and reach an underground detector, such that constraints from these underground experiments do not apply. Nevertheless, there are a number of direct detection experiments that have taken data on the surface of the Earth or even in space, which can potentially probe the same range of cross sections as the CMB~\cite{Xu:2018efh,Emken:2018run}. However, existing analyses make very specific assumptions on how the scattering rate scales for different target materials (i.e.\ that it is proportional to the nuclear mass squared), while the effects discussed here require no such assumption. Moreover, direct detection experiments are typically not sensitive to DM masses below a few hundred MeV, while CMB constraints on DM--baryon interactions remain valid down to the MeV-scale.\footnote{We note that it may be possible to extend the reach of direct detection experiments to lower masses by considering the non-detection of a sub-dominant component of DM particles that have been accelerated through collisions with cosmic rays~\cite{Bringmann:2018cvk}, but this approach also requires further model-dependent assumptions.} 
\enlargethispage{1\baselineskip}
For $n_b < 0$, on the other hand, the cross section grows with decreasing velocity and is, therefore, much larger at the time of recombination than in the present universe, where the typical DM velocities relevant for laboratory experiments are of the order of $10^{-3}c$. Whether this leads to a suppression of direct detection constraints (because the scattering rate is reduced) or an enhancement (because the stopping of DM particles in the Earth becomes negligible) is difficult to estimate in a model-independent way. A detailed comparison of cosmological constraints and direct detection experiments for these scenarios is, therefore, beyond the scope of the present work.

Finally, we note in passing that models with DM--baryon interactions would typically also feature DM--electron interactions. However, the cross section for the latter is expected to be suppressed proportional to $\mu_{\chi e}^2 / \mu_{\chi p}^2$, where $\mu$ denotes the reduced mass. In certain models, for example if the interactions arise from the exchange of a scalar mediator, even stronger suppression is possible. We therefore do not consider these interactions in the present work, even though they would be straightforward to implement in the formalism presented above, as explained in Ref.~\cite{Slatyer:2018aqg}.

\subsection{Dark matter -- photon interactions}\label{sec:th_dmg} 
Even though the defining property of DM is its lack of sizeable electromagnetic interactions, it is interesting to consider a sufficiently small but non-zero probability for DM--photon scattering. Indeed, such interactions would automatically be present in many of the models discussed above in the context of DM--baryon interactions, although they would be constrained to be very small. However, we emphasize that DM--photon interactions can also arise in models with suppressed DM--baryon interactions. For example, it has been pointed out that inelastic dipole transitions between different DM states may lead to effective DM--photon interactions at low energies that resemble Rayleigh scattering~\cite{Weiner:2012cb}.

In the present work, we focus on the case in which DM--photon interactions are independent of temperature,\footnote{Such a cross section arises, for example, if DM carries a fractional electric charge. Different types of interactions will, in general, lead to cross sections that decrease with decreasing temperature and, therefore, are less interesting in the present context.} and result in an additional term in the DM and photon velocity equations, analogous to the standard baryon--photon interaction term. The Euler equation for DM will thus be modified as
\begin{align}
\theta^\prime_\DM & = \theta^\prime_\mathrm{DM, standard} - \Gamma_{\DM\text{--}\gamma} \left(\theta_\DM -\theta_\gamma \right) \,,
\label{eq:addboltzdmg}
\end{align}
where 
\begin{equation}
\Gamma_{\DM\text{--}\gamma} = \frac{4\rho_\gamma} {3\rho_\DM} a \, \sigma_{\DM\text{--}\gamma}\,   n_\DM
\label{eq:gamma_dm_g}
\end{equation} 
is the {\it conformal DM--photon momentum exchange rate},
$\sigma_{\DM\text{--}\gamma}$ is the DM--photon elastic scattering cross section, and $n_\DM = \rho_\DM /m_\DM$ is the DM number density.
The other perturbation equations for baryons and photons are shown in full in App.~\ref{App:eqs}.
This case was already studied and implemented in \class in Refs.~\cite{Wilkinson:2013kia, Stadler:2018jin}. Coinciding with their paper release, the authors of~\cite{Stadler:2018jin} also released their own modified \class version.\footnote{\url{https://github.com/bufeo/class_v2.6_gcdm.git}} Our implementation is similar to theirs, but incorporated into our multi-interacting DM framework. Our code has been thoroughly cross-checked against theirs and produces the same results. 

Following Refs.~\cite{Wilkinson:2013kia, Stadler:2018jin}, it is convenient to define the scattering cross section relative to the Thompson cross section $\sigma_\mathrm{Th}$, and to introduce the dimensionless parameter
\begin{equation}
u_{\DM\text{--}\gamma} = \frac{\sigma_{\DM\text{--}\gamma}}{\sigma_\mathrm{Th}} \left( \frac{m_\DM}{100\,\mathrm{GeV}} \right)^{-1} \,,
\label{eq:dmg}
\end{equation}
such that
\begin{equation}
\sigma_{\DM\text{--}\gamma} = 6.65 \times 10^{-29}  u_{\DM\text{--}\gamma} \left( \frac{m_\DM}{100\,\mathrm{GeV}} \right) \mathrm{m}^2 \,.
\label{eq:gamma_dm_g2}
\end{equation} 
The rate $\Gamma_{\DM\text{--}\gamma}$ scales like $(1+z)^3$, while during radiation domination the conformal Hubble rate scales like $\H \propto (1+z)$: thus the DM--photon exchange rate is always more efficient in the early universe. 
The calculations of App.~\ref{App:dec} show that for models compatible with CMB bounds (that is, $u_{\DM\text{--}\gamma} \leq \mathcal{O}(10^{-4})$, as shown in the result section), DM always decouples from photons during radiation domination:
\begin{equation}
1+z_{\DM\text{--}\gamma} = 2.54\times10^{4} \left(\frac{u_{\DM\text{--}\gamma}}{10^{-4}}\right)^{-1/2} \,,
\end{equation}
while photons start evolving independently from DM even earlier (see equation~(\ref{eq:decgdm})).

We note that complementary constraints on DM--photon and DM--baryon interactions can be obtained from the halo mass function, which probes the non-linear matter power spectrum at small scales~\cite{Boehm:2014vja,Escudero:2018thh,Maamari:2020aqz}. Although these constraints can be stronger than the ones obtained from the CMB, they either require input from N-body simulations or analytical approximations. 
Finally, some strong bounds can be derived from the study of CMB spectral distortions and from FIRAS data, but such bounds only apply to a narrow range of DM masses, from about 1~keV to 100~keV \cite{Ali-Haimoud:2015pwa}.
Therefore, in the present work we do not include these effects and instead focus on the robust and model-independent constraints that can be obtained from the CMB alone. 

\subsection{Dark matter -- dark radiation interactions}\label{sec:th_dmdr} 

In analogy with the DM--photon interactions discussed above, DM can also interact with other forms of radiation. We consider the possibility that DM interacts with massless relics from the dark sector, called generically dark radiation (DR), which have negligible interactions with Standard Model particles. The general framework for such interactions has been developed in the ETHOS formalism~\cite{Cyr-Racine:2015ihg}, which also describes in detail the mapping between the underlying particle physics model and its effects on structure formation observables (see sections II A and II B of Ref.~\cite{Cyr-Racine:2015ihg}). The ETHOS parametrisation assumes that a single DM species interacts with a relativistic component via the 2-to-2 scattering $ \text{DM} + \text{DR}  \leftrightarrow \text{DM} + \text{DR}$. In addition we also include DR self-interactions via the process $ \text{DR} + \text{DR} \leftrightarrow \text{DR} + \text{DR}$, following the ETHOS implementation in \class from Refs.~\cite{Archidiacono:2017slj, Archidiacono:2019wdp}.

Note that DM could also interact with ordinary neutrinos. As long as neutrinos are approximated as massless, the formalism for DM--neutrino interactions can be seen as a sub-case of the ETHOS one, with a density of DR particles matched to the standard neutrino value $N_\mathrm{eff}\simeq3.044$ \cite{deSalas:2016ztq,Froustey:2020mcq, Akita:2020szl}, and assuming no self-interactions. Instead, interactions between DM and massive (active or sterile) neutrinos would require further extensions of our code.
\enlargethispage{2\baselineskip}

Within the ETHOS formalism, it is assumed that DR maintains a thermal spectrum with $T_\DR \propto (1+z)$ and vanishing chemical potential (such that $n_\DR \propto T_\DR^3$) throughout the times relevant for CMB physics and until today (see Ref.~\cite{Cyr-Racine:2015ihg} for a general discussion of these assumptions, or Ref.~\cite{Buen-Abad:2015ova} for a concrete example). These assumptions are consistent with the presence of {\it sufficiently weak} DM--DR interactions, such that any temperature drift or spectral distortion in the DR spectrum is negligible. On the other hand, since the number density of DM particles is much smaller than that of DR particles, the DM--DR interactions may have a significant impact on the evolution of the DM temperature, which we take into account in the full equations of App.~\ref{App:eqs}. With this formalism, the Euler equation for DM gains an additional term: 
\begin{equation}
\theta^\prime_\DM = \theta^\prime_\mathrm{DM, standard} - \Gamma_{\DM\text{--}\DR} \left( \theta_\DM - \theta_\DR \right) \,,
\label{eq:addboltzdmdr}
\end{equation}
where $\Gamma_{\DM\text{--}\DR}$ is the {\it conformal DM--DR momentum exchange rate}.
The DR perturbations are described by a Boltzmann hierarchy integrated over momentum, like in the case of massless neutrinos. When the DR self-interactions are assumed to be very strong, we truncate these equations at the level of the first two mutipoles, like for a relativistic perfect fluid.

We consider the case in which the interaction rate appearing in the DR equations has a power-law dependence on temperature
and can thus be written as 
\begin{equation}
\Gamma_{\DR\text{--}\DM}=\omega_\DM \, a_\mathrm{dark} \left(\frac{1+z}{1+z_d} \right)^{n_\DR} \, ,
\label{eq:coldrdm}
\end{equation}
where $\omega_\DM = \Omega_\mathrm{DM, 0}h^2$, while the rate
$a_\mathrm{dark}$ gives the overall interaction strength close to $z_d$, $n_\DR$ is the power-law dependence of the temperature, and $1 + z_d$ is a normalisation factor.\footnote{For models in which the DM and DR are in equilibrium at early times, it is convenient to pick $z_d$ close to the time of kinetic decoupling between the two species. The default value is $z_d=10^7$, corresponding to $T_\mathrm{kd}\sim1\,$keV.} The scattering rate for DM is given by
\begin{equation}
\Gamma_{\DM\text{--}\DR}=\left(\frac{4}{3} \frac{\rho_\DR}{\rho_\DM}\right) \Gamma_{\DR\text{--}\DM} \, ,
\label{eq:coldmdr}
\end{equation}
which is proportional to $(1+z)^{n_\DR+1}$, due to the different redshift dependence of $\rho_\text{DR}$ and $\rho_\text{DM}$. 
In principle, one can also calculate the self-scattering rate $\Gamma_{\DR\text{--}\DR}$ for a given model, but in the case of strong self-coupling, DR behaves like a perfect fluid and the precise value of $\Gamma_{\DR\text{--}\DR}$ becomes irrelevant.

Once the pivot redshift $z_d$ of equation (\ref{eq:coldrdm}) has been fixed arbitarily to $z_d=10^7$, like in previous works \cite{Cyr-Racine:2015ihg,Archidiacono:2017slj,Archidiacono:2019wdp}, 
the conformal DM--DR momentum exchange rate (\ref{eq:coldrdm}) can be conveniently parametrised either in terms of $(a_\mathrm{dark} a_0^{-1})$ or of the current rate
\begin{equation}
\Gamma_{\DM\text{--}\DR}^0 \equiv \Gamma_{\DM\text{--}\DR}(z=0) \, a_0^{-1} =\frac{4}{3} \omega_\mathrm{\DR} a_\mathrm{dark} \, a_0^{-1} 10^{-7 n_\DR} \, ,
\end{equation}
with $\omega_\DR = \Omega_\mathrm{DR, 0}h^2$, such that
\begin{equation}
	\Gamma_{\DM\text{--}\DR} = \frac{4}{3} \omega_\mathrm{\DR} a_\mathrm{dark} (1+z) \left(\frac{1+z}{10^7}\right)^{n_\DR} = \Gamma_{\DM\text{--}\DR}^0 a_0 \left({1+z}\right)^{1+n_\DR} \, .
	\label{eq:gammadmdrpar}
\end{equation}
In the case of $n_\DR = \{2,4\}$, Refs.~\cite{Archidiacono:2017slj,Archidiacono:2019wdp} report their observational bounds on the parameter $a_\mathrm{dark}$ (assuming $a_0=1$ and $z_d=10^7$). For $n_\DR=0$, Refs.~\mbox{\cite{Lesgourgues:2015wza,Buen-Abad:2017gxg,Archidiacono:2019wdp}} report bounds on $\Gamma_{\DM\text{--}\DR}^0$, which is just denoted by $\Gamma_0$ in these works. The parameters $\Gamma_{\DM\text{--}\DR}^0$ and $a_\mathrm{dark}$ have the dimension of rates, but their bounds are often expressed in inverse Megaparsecs (using $c=a_0=1$). The correspondence with inverse seconds is given by
\begin{equation}
{1.029 \, \mathrm{Mpc}^{-1}} \simeq {10^{-14} \mathrm{s}^{-1}}\,.
\end{equation}
For $n_\DR>0$, the DM--DR momentum exchange rate is more efficient in the early universe, and for $n_\DR<0$ in the late universe. The limiting case $n_\DR=0$ corresponds to a rate scaling as $\Gamma_{\DM\text{--}\DR} \propto (1+z)$, while during radiation domination, $\H$ also scales like $(1+z)$: thus, the influence of the DM--DR interactions can remain small but constant throughout this stage.

For $n_\DR > 0$, by equating the rate in equation (\ref{eq:gammadmdrpar}) with the conformal Hubble rate $\H$ during radiation domination, one gets an approximation for the redshift at which DM decouples from DR (see App.~\ref{App:dec}, equation~(\ref{eq:dmdrdec_app}) for further details):
\begin{equation}
1+z_{\DM\text{--}\DR} \sim 10^{8+\frac{2}{n_\DR}}
 \left(
\frac{\Gamma_{\DM\text{--}\DR}^0}{10^{-8n_\DR-22} \, \mathrm{s}^{-1}}
\right)^{-\frac{1}{n_\DR}}\,.
\label{eq:dmdrdec}
\end{equation}
According to Ref.~\cite{Archidiacono:2019wdp}, typical bounds from CMB and Lyman-$\alpha$ data are roughly of the order\footnote{Indeed, Ref.~\cite{Archidiacono:2019wdp} finds that the bounds are described in all cases by $10^{4-n_\DR} (a_\mathrm{dark}/\mathrm{Mpc}^{-1})\xi^4 < {\cal O}(10)$, where $\xi$ is the DR-to-photon temperature ratio, such that $\omega_\DR=\xi^4 \omega_\gamma$ (with possibly one extra factor $\frac{7}{8}$ for fermions). This gives a bound $a_\mathrm{dark} \omega_\DR < {\cal O}(10^{-n_\DR-3}) \, \omega_\gamma \mathrm{Mpc}^{-1}$, which can be turned into $\Gamma_{\DM\text{--}\DR}^0 < {\cal O} (10^{-8 n_\DR-22}) \, \mathrm{s}^{-1}$.} of $\Gamma_{\DM\text{--}\DR}^0 < {\cal O} (10^{-8 n_\DR-22}) \, \mathrm{s}^{-1}$. Thus, for typical values of $\Gamma_{\DM\text{--}\DR}^0$ compatible with observations and or $n_\DR > 0$, equation (\ref{eq:dmdrdec}) shows that DM--DR decoupling takes place during radiation domination. 

Since our code is based on the previous work of Refs.~\cite{Archidiacono:2017slj, Archidiacono:2019wdp} (extended to take into account simultaneous interactions), it covers the general ETHOS case, and hence various values of $n_\DR$ and  several possible assumptions concerning the DR self-interaction rate. However, in the comparison with observations presented below, we will focus specifically on a model known to be particularly relevant for the discussion of the Hubble and $S_8$ tensions
\cite{Lesgourgues:2015wza, Buen-Abad:2017gxg,Archidiacono:2019wdp}. In this model, one chooses $n_\DR=0$ such that  $\Gamma_{\DM\text{--}\DR}/\H$  remains constant throughout radiation domination, and decreases during matter domination. Then, the small but cumulative effect of DM--DR scattering throughout radiation domination can lead to a small enhancement of DR fluctuations and to a small suppression of DM  fluctuations that have interesting consequences for the CMB and matter power spectra. In this model, one
further assumes that DM--DR interactions are too weak to bring the two species into thermal equilibrium, while DR has strong self-interactions and behaves as a perfect fluid (not free-streaming). 

This class of models is easy to motivate with a concrete dark sector set-up, like for instance in the non-Abellian Dark Matter model of Ref.~\cite{Buen-Abad:2015ova}. It can be described by two parameters
\begin{equation}
(\Gamma_{\DM\text{--}\DR}^0, \, \Delta N_\DR)\, ,
\end{equation}
where $\Delta N_\DR \equiv \frac{\rho_\DR}{\rho_{1\nu}}$ gives the amount of DR relative to the energy density of a single neutrino species in the instantaneous decoupling approximation. We emphasize that for a given model of DR, this parameter also fixes the DR temperature $T_\DR$\,. In the following, we will consider the case that DR has two bosonic degrees of freedom, which implies
\begin{equation}
 \Delta N_\text{DR} \approx 8.8 \times \left(\frac{T_\DR}{T_\gamma}\right)^4 \, .
\end{equation}
For the parameters that we will consider, the DR does not thermalise with either photons or DM and, therefore, $\Delta N_\text{DR}$ (or equivalently $T_\DR$) is a free parameter.\footnote{If, on the other hand, DR was in thermal equilibrium with photons above some high temperature $T_\text{dec}$, the photon-to-DR temperature ratio would be dictated by entropy conservation. For instance, if DR has two bosonic degrees of freedom, one finds $
 \Delta N_\text{DR} \approx 8.8 \times \left(\frac{g_\ast}{g_\text{dec}}\right)^{4/3} \, ,$
where $g_\ast$ denotes the number of effective relativistic degrees of freedom and $g_\text{dec} = g_\ast(T = T_\text{dec})$. 
For instance, assuming $g_\text{dec} \sim 90$ (corresponding roughly to $T_\text{dec} \sim 10\,\mathrm{GeV}$), one gets $\Delta N_\text{DR} \sim 0.07$ during recombination.
In previous analyses, this emblematic value was sometimes assumed as a lower bound on the prior of $N_\text{DR}$. In the present work, we do not impose such a bound and instead remain agnostic about the value of $ \Delta N_\text{DR}$.} 

\subsection{Temperature evolution of multi-interacting dark matter}\label{sec:th_idm} 

In the case of multi-interacting DM, most of the ingredients described previously can be aggregated in a straightforward manner, as can be seen in the full equations presented in App.~\ref{App:eqs}. However, a few aspects require special attention when more than one DM scattering channel is turned on.

First, in the combined interaction model, the DM temperature needs to be calculated consistently and evolved together with the baryon temperature. This is not needed for photons and DR particles, for which the assumptions described in the previous sections imply that their temperature scales as $T\propto (1+z)$. The value of the DM temperature is relevant for the calculation of the DM--baryon momentum exchange rate given by equation~(\ref{eq:coldmb}), and for that of the DM sound speed appearing in the DM Euler equation. We will come back to the relevance of the sound speed at the end of this section.

The evolution equation for the DM temperature depends on all interaction rates,\footnote{For DM--baryon, we assumed here for simplicity that DM only scatters off hydrogen atoms, which leads to ${\cal F}_\mathrm{He}=1-Y_p$ (and thus to $R_\chi=R'_\chi$ in the notations of Ref.~\cite{Dvorkin:2013cea}, see their equation (15)). This leads to the appearance of $\Gamma_{\DM\text{--}b}$ without Helium correction terms in the equation.}
\begin{align}
{T}_{\DM}^\prime = -2\H T_{\DM} &- 2  \Gamma_{\DM\text{--}\gamma} (T_\DM - T_{\g}) \nonumber \\
& - 2 \Gamma_{\DM\text{--}\DR} (T_\DM - T_{\DR}) \nonumber \\
&- \frac{2m_{\DM}}{m_{\DM} + m_{b} } \Gamma_{\DM\text{--}b}  (T_\DM - T_{b}) \, .
\label{eq:Tdm}
\end{align}
The rates in front of each term $(T_\DM - T_x)$ are the \textit{ conformal heat exchange rates} between DM and each species $x$. They are related to the respective momentum exchange rates, because they are derived from the same collision operator in the Boltzmann equation. Assuming that each scatter changes the momentum of the non-relativistic DM particle only by a small amount, one can analytically derive\footnote{As shown in Ref.~\cite{Bringmann:2016ilk} for the case of DM--DR scattering, equation~(A67) of \cite{Cyr-Racine:2015ihg} can be written as  $\Gamma^\mathrm{momentum}_{\DM\text{--DR}} = \frac{\eta_\DM}{6\pi^2 m_\DM} \int \mathrm{d}\omega g^{\pm}(\omega) \partial_\omega (\omega^4 \sigma_T)$, where $\eta_\text{DM}$ denotes the $\DM$ spin degrees of freedom, $\omega$ is the energy of the DR and $g^\pm(\omega)$ the DR thermal distribution. This expression is the same as equation~(A9) of \cite{Bringmann:2016ilk} up to a factor of 2, which establishes the link for massless DR particles (see also appendix B of \cite{Bringmann:2016ilk}). It is easy to generalise this result to massive DR particles as well, provided their distribution remains thermal.} $\Gamma^\mathrm{heat}_{\DM\text{--}x} = 2\Gamma^\mathrm{momentum}_{\DM\text{--}x}$ for scattering with a massless species, and a similar relation with additional mass factors for non-relativistic scattering partners (such as baryons). 
An explicit calculation for the ETHOS $n_\DR=0$ case in the context of non-Abellian Dark Matter is provided in Ref.~\cite{Buen-Abad:2015ova}.

To follow the temperature evolution of equation~(\ref{eq:Tdm}), we need to impose initial conditions for the DM temperature at the earliest time considered by the \class thermodynamics module.
By default, this time would correspond to the redshift $z_\mathrm{ini}=5 \cdot 10^6$, but in the presence of IDM the \class thermodynamics module starts earlier, as described below.
\newpage
\noindent To start from a plausible initial value of $T_\DM$, we impose the following conditions:
\begin{enumerate}[leftmargin=0.5cm]
	\item In the models where there is a strong coupling at early times (such as DM--photon, DM--baryon with $n_b > -3$, or DM--DR $n_\DR > 0$) we use the analytic approximations of the decoupling redshifts listed in previous sections (and derived in App.~\ref{App:dec}). The latest decoupling redshift $z_\text{dec}$ is used to determine the starting point of integration, which is taken to be $z_\text{ini} = 10^4z_\text{dec}$\,. At these times the coupling is definitely strong enough ($\Gamma_{\DM\text{--}x} \gg \H$) to justify tightly coupled initial conditions.
	
	\begin{enumerate}[leftmargin=0.65cm]
		\item For DM--photon and DM--baryon couplings the initial temperature is chosen to be $T_\DM=T_b=T_\gamma=T_\gamma^0(1+z)$, where $T_\gamma^0$ is the present-day CMB temperature of 2.7255K \cite{Fixsen:2009ug}.
		
		\item For the DM--DR interactions we set instead $T_\DM=T_\DR =T_\DR^0(1+z)$. While DM and DR might have been in thermal equilibrium with the visible sector at some even earlier time, we do not have to set $T_\DR=T_\gamma$, since the coupling to the visible sector is assumed to be small at $z_\mathrm{ini}$\, for this case. In principle, we expect that after the decoupling between the visible and the dark sector, the ratio of visible-to-dark temperatures will have been enhanced by entropy releases in the visible sector and/or decreased by entropy releases in the dark sector. Thus we can consider $T_\DR^0$ as a free parameter.
		
		\item If DM is initially in thermal equilibrium with the visible sector (through DM--photon or DM--baryon interactions with $n_b = \{-2, 0\}$) and also with DR through DM--DR interactions with $n_\DR>0$, our code can handle it, but for self-consistency, the user should then choose $T_\DR=T_\gamma$ at initial times, or a value of $\Delta N_\DR$ compatible with this assumption. Moreover, our code detects this situation and raises an error if the user chooses $T_\DR \neq T_\gamma$. Note that this case is not overly interesting, because it leads to values of $\Delta N_\DR$ of order one, which are likely to conflict with BBN and CMB bounds on $N_\mathrm{eff}$ \cite{Aghanim:2018eyx,Schoneberg:2019wmt}. 
	\end{enumerate} 

	\item In those cases where there is no strong coupling at early times, the starting redshift of integration is set to $z_\text{ini} = 10^8$. 
	\begin{enumerate}[leftmargin=0.65cm]
		\item If there are DM--DR interactions with $n_\DR=0$, the ratio $\epsilon\equiv2\Gamma_{\DM\text{--}\DR}/\H$ is constant throughout radiation domination (see App.~\ref{App:dec}). This ratio is given by equation~\eqref{eq:dmdrdec_app3} and needs to be much smaller than one for models compatible with the data. Thus DM and DR interact too weakly at $z_\mathrm{ini}$ to be in thermal equilibrium.
		Nevertheless, their small interaction rate implies that $T_\DM$ is driven towards a steady-state attractor solution where $T_\DM= \frac{\epsilon}{(1+\epsilon)} T_\DR$. The thermodynamics module imposes such a condition at $z_\mathrm{ini}=10^8$.
		Here again due to entropy releases after the decoupling between the visible and the dark sector, we can consider $T_\DR^0$ as a free parameter. 
	
		\item Otherwise, that is, in the case of only DM--baryon interactions with $n_b = {-4}$, we assume that DM was either never in thermal equilibrium with photons at high temperatures or became much colder than photons due to several entropy releases after DM decoupling, and like previous authors (see e.g. Ref.~\cite{Slatyer:2018aqg}) we start from a null value of $T_\DM$ at~${z_\mathrm{ini}=10^{8}}$. 
	\end{enumerate}
\end{enumerate}

\begin{figure}[t]
	\centering
	\begin{subfigure}[b]{0.5\textwidth}
		\centering
		\includegraphics[width=\textwidth]{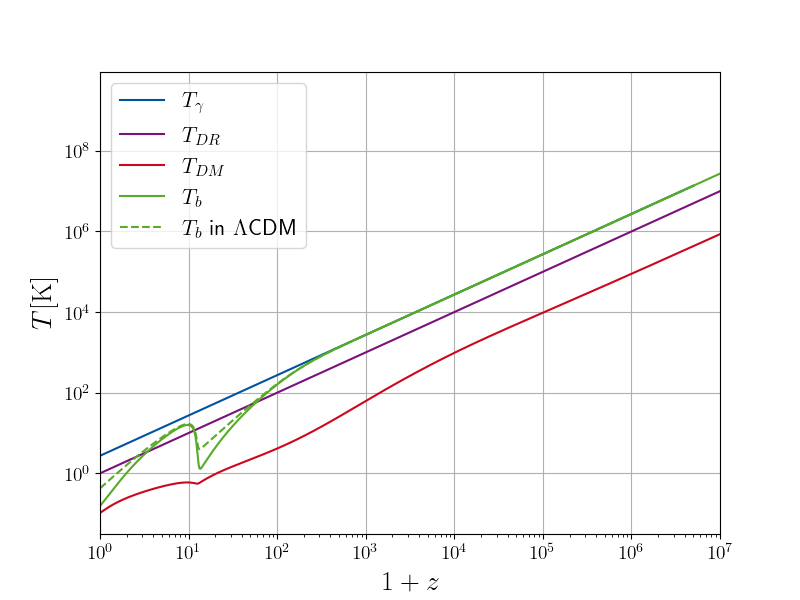}
	\end{subfigure}\hspace*{2mm}\begin{subfigure}[b]{0.5\textwidth}
		\centering
		\includegraphics[width=\textwidth]{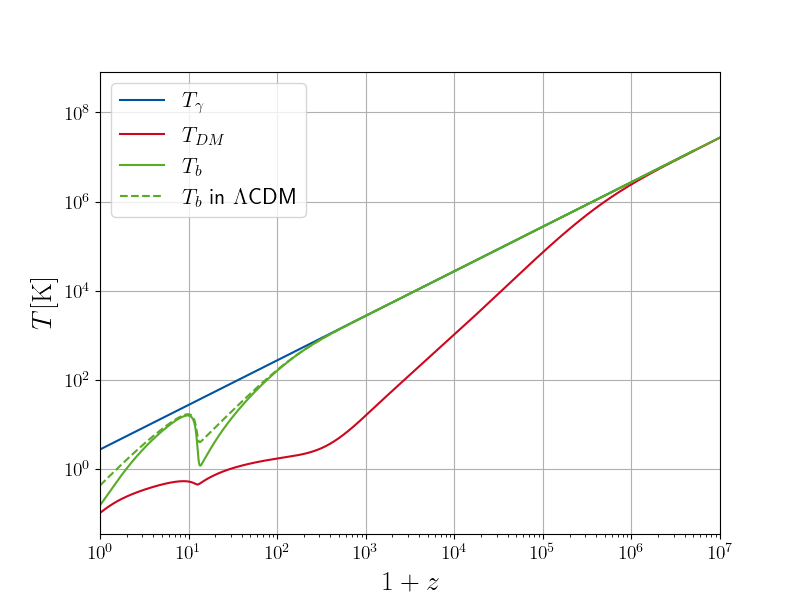}
	\end{subfigure}
	\caption{Temperature evolution for two examples of interacting models. \textbf{Left:} DM--DR and DM--baryon ($n_b=-4$) interactions with $N_\DR = 0.07$, $\Gamma^0_{\DM\text{--}\DR} = 10^{-7}$Mpc$^{-1}$, and $\sigma_{\DM\text{--}b} = 10^{-41}$cm$^2$. \textbf{Right:} DM--photon and DM--baryon ($n_b=-4$) interactions with $u_{\DM\text{--}\g} = 10^{-6}$ and $\sigma_{\DM\text{--}b} = 10^{-41}$cm$^2$. 
	} 
	\label{fig:temperature_evolution}
\end{figure}

\newpage
We show in figure \ref{fig:temperature_evolution} a few examples of models featuring a non-trivial DM temperature evolution. 
The left panel corresponds to a model with DM--DR interactions ($n_\DR=0$, $\Gamma_{\DM\text{--}\DR}^0=10^{-7} \mathrm{Mpc}^{-1}$, $\Delta N_\DR=0.07$)  plus DM--baryon interactions ($n_b=-4$, $\sigma_{\DM\text{--}b}=10^{-41} \mathrm{cm}^{2}$) for $m_\DM=1$~GeV. These values are (at least marginally) compatible with the bounds found in Sec.~\ref{sec:res}. Initially, the DM temperature is influenced only by DM--DR scattering, and follows the attractor solution $T_\DM= \frac{\epsilon}{(1+\epsilon)} T_\DR \propto (1+z)$ with $\epsilon\equiv2\Gamma_{\DM\text{--}\DR}/\H \approx 0.092$. 
After radiation-to-matter equality, DM gradually decouples and its temperature begins to drop faster, since without interaction it would cool adiabatically as $T_{\DM}\propto (1+z)^2$. However, before reaching the adiabatic behaviour, DM starts to feel the interaction with baryons and is consequently heated. 
We see that at the current epoch, $T_{\DM}$ is of the same order of magnitude as $T_b$. For the same model, we show the evolution of the baryon temperature taking into account the DM--baryon heating rate (green curve) or neglecting it (green dashed curve). When the interaction is neglected, the baryons start to cool adiabatically after their decoupling from photons; then, they get reheated by reionization; and finally, they cool again adiabatically. Note that \class models the effect of reionization on the baryon temperature in a very approximate way, to be improved in future versions. When the interactions are taken into account, we can see that the baryons are further cooled at late times, because they transfer heat to DM.

The right panel of figure \ref{fig:temperature_evolution} features DM interacting with photons ($u_{\DM\text{--}\gamma} =10^{-6}$) and baryons ($n_b=-4$, $\sigma_{\DM\text{--}b}=10^{-41}$) for $m_\DM=1$~GeV. These values are well within the bounds found in the result section (Sec.~\ref{sec:res}). The DM temperature initially follows the photon one, until DM decouples from photons around $z\sim{\cal O}(10^6)$. 
Then, DM cools adiabatically, until late interactions with baryons heat it again. The evolution of the baryon temperature is similar to that in the previous model.

\newpage
Once the DM temperature evolution is known, we can obtain its equation of state parameter $w_\DM=p_\DM/\rho_\DM$ and sound speed $c_\DM^2=\delta p_\DM/\delta \rho_\DM$ from
\begin{equation}
w_\DM =  \frac{k_B T_{\DM}}{m_{\DM}}\, , \qquad c_{\DM}^2 = \frac{k_B T_{\DM}}{m_{\DM}} \left(1 - \frac{1}{3} \pdv{\ln T_{\DM}}{\ln a} \right) \,,
\end{equation}
which are valid to first order in $T_\DM/m_\DM$. In general, the continuity and Euler equations governing the evolution of the perturbations of a given species depend on $w_\DM$ and on the sound speed, which appear in several terms. For ordinary decoupled CDM, these parameters are so small that they are totally neglected in Boltzmann codes (they would only affect the evolution of extremely small wavelengths crossing the Hubble radius extremely early).
In the models considered here, however, DM can have its temperature considerably enhanced through scattering. Thus, one could object that we need to take into account the non-zero value of $w_\DM$ and $c_{\DM}^2$ wherever necessary in the continuity and Euler equations. 
Nonetheless, we can derive an upper bound on $w_\DM$ and $c_\DM^2$ at the time when a given mode $k$ crosses the Hubble radius ($k=aH$) during radiation domination. 
We should keep in mind that the DM couplings discussed in this work can raise the DM temperature at most up to the photon temperature,\footnote{We only need to consider early times here where $T_\DM \leq T_b = T_\gamma$. While the late time IGM temperature can reach up to $\mathcal{O}(10^5)$K during reionization, even higher baryon temperatures are easily reached for $z\sim \mathcal{O}(10^5)$. The most stringent constraints then come from $z \gg \mathcal{O}(10^5)$.} 
such that $w_\DM \leq T_\gamma/m_\DM$ with $T_\gamma \propto (1+z)$, and $c_\DM^2 \leq \frac{4}{3} T_\gamma/m_\DM$. Then, we obtain the following upper bound at the time when $k=aH$: 
\begin{equation}
w_\DM, \, c_\DM^2  \leq 10^{-3} \left(\frac{1\,  \mathrm{MeV}}{m_\DM}\right) \left(\frac{k}{1 \, \mathrm{Mpc}^{-1}}\right)\,.
\end{equation}
This shows that even if the various couplings bring the DM up to the photon temperature, as long as we consider masses $m_\DM \gtrsim1 \, \mathrm{MeV}$ and wavenumbers $k \lesssim 1 \, \mathrm{Mpc}^{-1}$, the DM will always have $w_\DM<10^{-3}$ and $c_\DM^2 < 10^{-3}$ when the modes are in the sub-Hubble regime. Thus, as long as we assume $m_\DM \gtrsim 1 \, \mathrm{MeV}$, we do not need to include $w_\DM$ and $c_\DM^2$ everywhere in the continuity and Euler equations for computing the CMB spectra. Even if we are interested in Lyman-$\alpha$ physics and in $k\sim 50 \, \mathrm{Mpc}^{-1}$, our approximation holds at least as long as  $m_\DM \gtrsim 50 \, \mathrm{MeV}$.

However, following Ref.~\cite{Stadler:2018jin}, we do include the sound speed in the pressure source term of the Euler equation, $k^2 c_\DM^2 \delta_\DM$, where it could play a role for large enough values of $k$. In principle, this pressure term can affect the dynamics of the modes when $k^2 c_\DM^2 > \H^2$. The sound speed $c_\DM^2$ is largest when DM is strongly coupled during radiation domination to the visible sector, with $T_\DM \sim T_\gamma$. Then $k^2 c_\DM^2\propto (1+z)$ while $\H^2 \propto (1+z)^2$.
Thus the impact of $c_\DM^2$ is maximal near the time when the DM temperature decouples at some redshift $z_\DM^\mathrm{dec}$\,. At this redshift, we can estimate the ratio
\begin{equation}
\frac{k^2 c_\DM^2}{\H^2} \sim 6.7 \times 10^{-3} \left(\frac{1 \, \mathrm{MeV}}{m_\DM}\right) \left(\frac{10^4}{1+z_\DM^\mathrm{dec}}\right) \left(\frac{k}{1 \, \mathrm{Mpc}^{-1}}\right)^2 \, .
\end{equation}
This means that for any DM candidate with $m_\DM \gtrsim 1 \, \mathrm{MeV}$ decoupling at $z_\mathrm{dec} \gtrsim 10^4$, and for all modes $k \lesssim 1 \, \mathrm{Mpc}^{-1}$, the pressure term in the Euler equation is such that ${k^2 c_\DM^2} \ll \H^2$: thus we expect the impact of the sound speed on cosmological scales to be negligible. Indeed, assuming $m_\DM \sim 1 \, \mathrm{GeV}$, we checked that even for models saturating the bounds found in the results section on all types of interactions, the influence of the pressure term on the matter power spectrum at $k<10 \, \mathrm{Mpc}^{-1}$ is always negligible. We leave the pressure term in the code only for the sake of completeness.

Finally, we see that following the DM temperature $T_\DM(z)$ is mainly useful for getting a correct estimate of the DM--baryon momentum exchange rate following equation~(\ref{eq:coldmb}) even in the presence of other interactions such as DM--photons or DM--DR. This is important for computing CMB observables and matter power spectra, since it impacts the evolution of the matter and baryon density fluctuations $(\delta_\DM, \delta_b)$.  It is also important for following the evolution of  $T_b(z)$, and thus potentially for using observations of the IGM temperature and ionization fraction, of the 21cm differential brightness temperature, of the Sunyaev-Zeldovitch effect, or of CMB spectral distortions.

\subsection{Tight-coupling approximations with multi-interacting dark matter}\label{sec:th_tca} 

The models described here feature multiple possible combinations of tight-coupling regimes between photons, baryons, DM, and DR.
Whenever two or more species are tightly coupled, the system of perturbation equations becomes stiff. Fortunately, \class is using by default an implict ODE solver, \texttt{ndf15}, which is ideal for solving stiff systems \cite{Blas:2011rf}. Such an ingredient is crucial in the context of this work, because otherwise we would need to implement a complicated set of Tight-Coupling Approximations (TCAs) describing fifteen possible tight-coupling regimes between two, three, or four species.

However, there is a limit to the degree of stiffness that \texttt{ndf15} can handle. Thus, when a scattering rate exceeds the Hubble rate by many orders of magnitude, it is still advisable to switch from the exact equations to TCA equations, which are derived from a perturbative expansion of the solution of the equation for the differential velocity $(\theta_x-\theta_y)$ in the inverse scattering rate \cite{Ma:1995ey,Blas:2011rf}. 

For DM--DR interactions with $n_\DR=4$ (or even larger), such an ``extreme tight-coupling regime''  can be reached at the earliest times considered by the Boltzmann code, because in this case $\Gamma_{\DM\text{--}\DR}/\H$ scales initially like $(1+z)^3$, instead of $(1+z)^2$ for Thomson baryon--photon scattering and DM--photon scattering, or $(1+z)$ for DM--DR scattering with $n_\DR=2$. Thus, the authors of Ref.~\cite{Archidiacono:2017slj} developed and implemented in \class a DM--DR TCA, which is used at early times and switched off automatically when $\Gamma_{\DM\text{--}\DR}/\H$ drops below a given threshold. Since the present work is an extension of Ref.~\cite{Archidiacono:2017slj}, our code still includes the DM--DR TCA scheme, even though it is not needed for the $n_\DR=0$ case considered in the next sections.

The photon--baryon TCA scheme is less essential to \class when using \texttt{ndf15}, but it has been implemented in the code since the beginning, and it does improve its performance. Thus, our code features two TCAs: one for the visible sector, and one for the dark sector.

We have seen that our models for DM--baryon and DM--photon interactions imply that the visible sector and at least part of the dark sector can be tightly coupled until the end of radiation domination (for DM--photon and DM--baryon with $n_b>-3$), or at later times (in the case of DM--baryon with $n_b=-4$). Fortunately, CMB bounds are such that \class only needs to deal with a moderate degree of stiffness for these interactions, that can be perfectly handled by \texttt{ndf15} without requiring further TCAs. Indeed, the code is not significantly slowed down during the new tight-coupling epochs, and the solution of the perturbation equations remains smooth and well-converged.

\begin{figure}[t]
	\centering
	\begin{subfigure}{0.48\linewidth}
		\centering
		\includegraphics[width=1\textwidth]{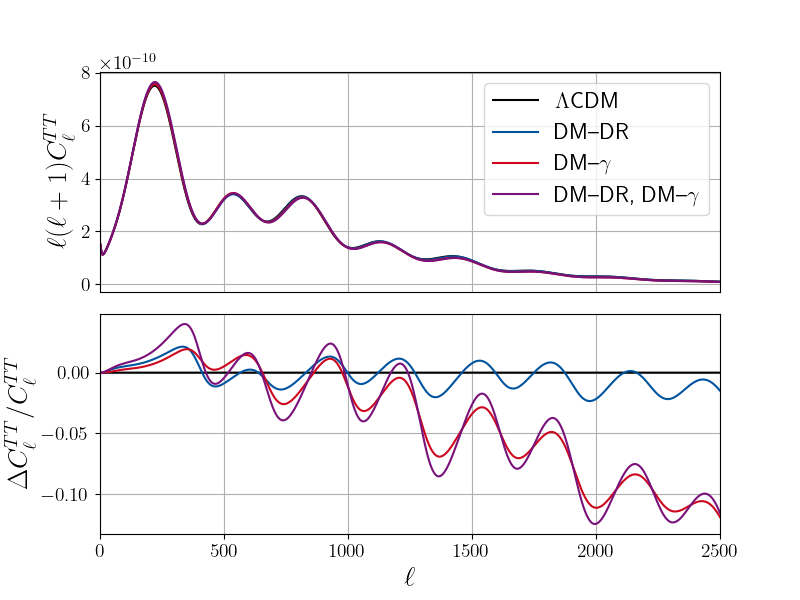}
	\end{subfigure}%
	\begin{subfigure}{0.48\linewidth}
		\centering
		\includegraphics[width=1\textwidth]{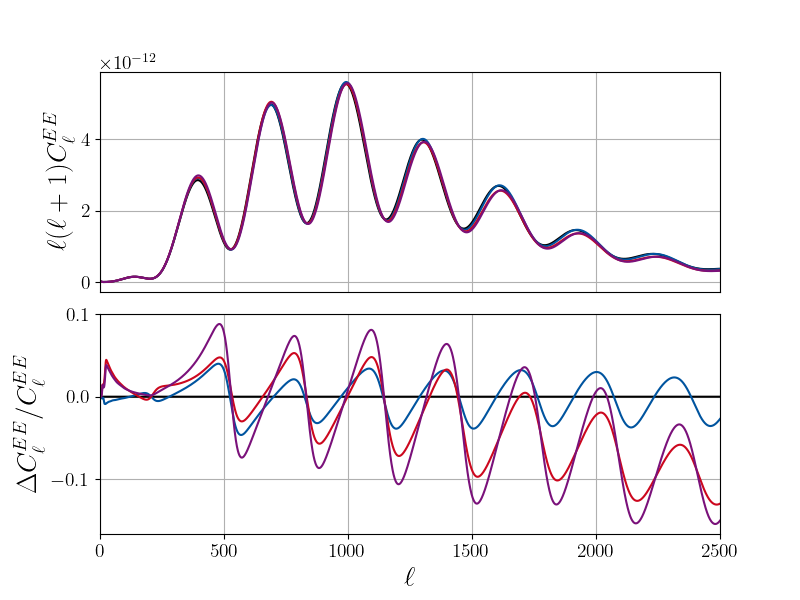}
	\end{subfigure}\\
	\begin{subfigure}{0.48\linewidth}
		\centering
		\includegraphics[width=1\textwidth]{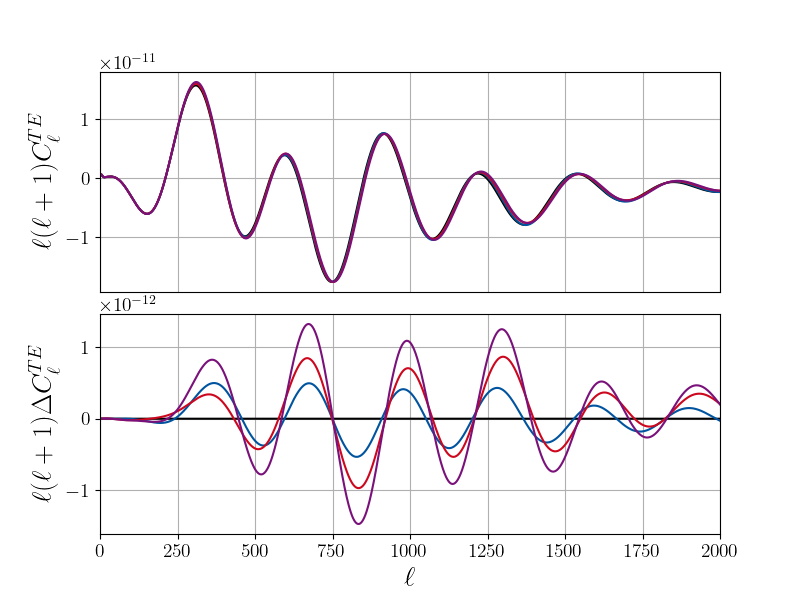}
	\end{subfigure}%
	\begin{subfigure}{0.48\linewidth}
		\centering
		\includegraphics[width=1\textwidth]{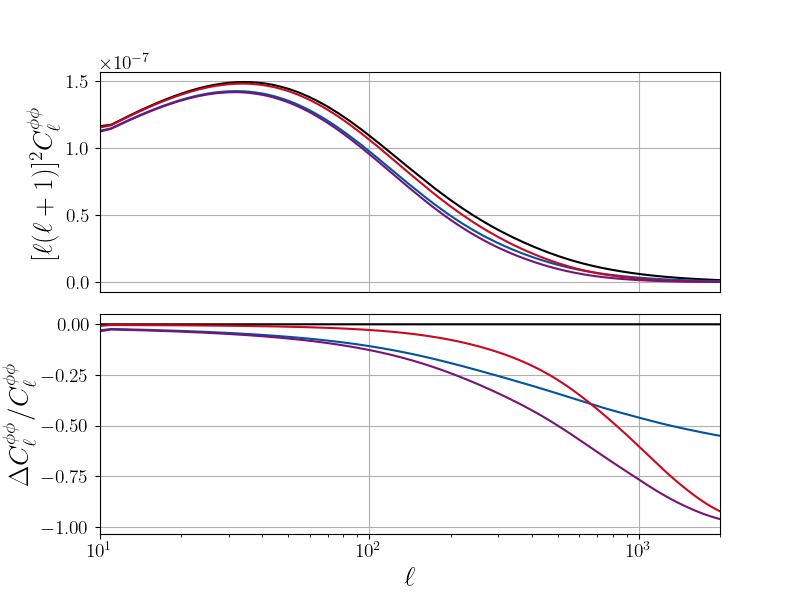}
	\end{subfigure}
	\caption{Effect of (individual or combined) DM--DR and DM--photon interactions on the CMB anisotropy spectra. \textbf{Top left:} Temperature spectra and relative residuals. \textbf{Top right:} E-mode polarisation spectra and relative residuals.\textbf{Bottom left:} Temperature-polarisation cross correlation spectra and absolute residuals. \textbf{Bottom right:} Lensing deflection spectra and relative residuals.
}
	\label{fig:idm_g_dr_cl}
\end{figure}

Nevertheless, each of the DM interactions must be taken into account while solving the photon--baryon TCA equations: even during this epoch, photons and/or baryons can be influenced by the scattering with DM. The generalisation of the photon--baryon TCA equations to incorporate DM--baryon (resp. DM--photon) scattering was already presented in Ref.~\cite{Xu:2018efh} (resp. Ref.~\cite{Stadler:2018jin}). We generalised it further to the case of multi-interacting DM. Due to the complexity of the problem, our calculation is based on the first-order TCA scheme called \texttt{first\_order\_CLASS} rather than the default scheme (which also contains the less suppressed order-two terms and is called \texttt{compromise\_CLASS}): we checked that this has a negligible impact on the final precision of the code. 
Similarly, we modified the DM--DR TCA equations in order to take into account the influence of DM--photon and/or DM--baryon scattering. 
Our modified TCA equations are summarised in App.~\ref{App:eqs}. 

\subsection{Impact of multi-interacting dark matter}\label{sec:th_imp}
The impact of each single DM scattering channel has been described in several previous works, already mentioned in the introduction section. Using our multi-interaction code, we find empirically that these effects tend to sum up in a rather straightforward manner, such that the effects of dual or triple interactions are very similar to the summed effects from each channel. This can be seen at the level of the CMB and matter power spectra for individual models, and will be further confirmed by the confidence limits derived in Secs.~\ref{sec:res} and~\ref{sec:tensions}.

\begin{figure}[t]
	\centering
	\includegraphics[width=0.55\textwidth]{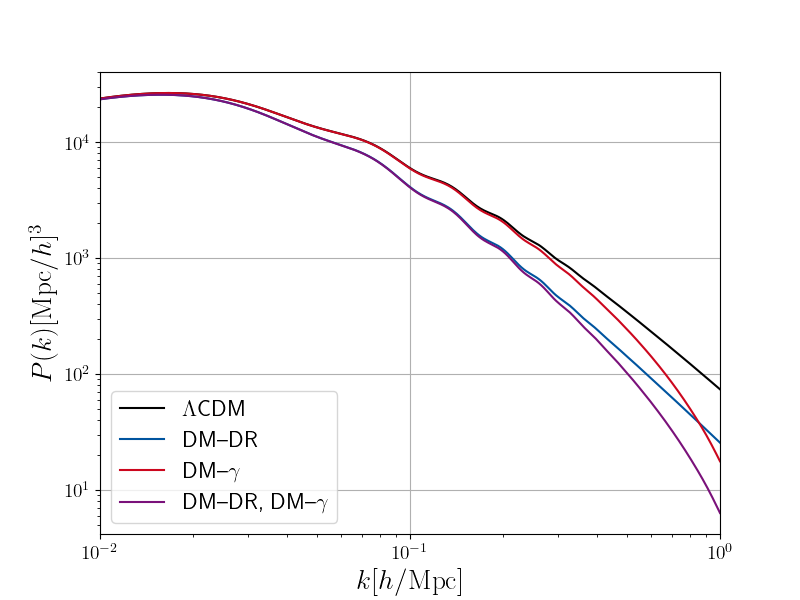}
	\caption{Effect of (individual or combined) DM--DR and DM--photon interactions on the matter power spectrum.	}
	\label{fig:idm_g_dr_pk}
	\vspace*{-1\baselineskip}
\end{figure}

\vspace*{-0.25\baselineskip}
We illustrate this additive trend in figures \ref{fig:idm_g_dr_cl} and \ref{fig:idm_g_dr_pk}, for the particular example of DM interacting simultaneously with photons and DR ($n_\DR=0$). For these figures, we have assumed $m_\DM=1 \, \mathrm{GeV}$, $N_\DR = 0.07$, $\Gamma^0_{\DM\text{--}\DR} = 5\times10^{-7}$Mpc$^{-1}$, and  $u_{\DM\text{--}\g} = 10^{-3}$.

\vspace*{-0.25\baselineskip}
The DM--photon interactions are known to have the following effects: suppress the small-scale CMB spectra due to collisional damping, shift the peaks to smaller scales due to a reduction of the sound speed, and suppress the small-scale matter power spectrum exponentially due to the DM being dragged by the photons \cite{Boehm:2004th,Wilkinson:2013kia,Stadler:2018jin}. These effects are clearly visible when comparing the black and red curves in figures \ref{fig:idm_g_dr_cl} and \ref{fig:idm_g_dr_pk}. 

\vspace*{-0.25\baselineskip}
The DM--DR interactions with $n_\DR=0$ have a smaller effect on the CMB. Normally, extra free-streaming radiation suppresses the small-scale CMB spectrum due to Silk damping and shifts the acoustic peaks due to neutrino drag. These effects are much smaller with the DR component of the $n_\DR=0$ model, because small-scale photon perturbations are also boosted by the DR perturbations, which are larger than those of free-streaming neutrinos due to the DM--DR scattering. The DR component also has a smaller sound speed due to its self-interactions. Furthermore, the $n_\DR=0$ model is also known for suppressing the matter power spectrum in a special way, due to DM being dragged by DR over the radiation dominated epoch. The suppression is smoother and affects larger scales than with other interacting DM models \cite{Lesgourgues:2015wza,Buen-Abad:2017gxg}.
These effects can also be seen in the blue curves in figures \ref{fig:idm_g_dr_cl} and \ref{fig:idm_g_dr_pk}. 
\enlargethispage{3\baselineskip}

\vspace*{-0.25\baselineskip}
Finally, in all panels, one can check that the combined effect of simultaneous DM--DR and DM--photon interactions (purple curves) looks qualitatively very similar to the sum of the individual effects, showing that these effects are largely additive. We leave for future work a discussion of the potential of these models to address the $A_\mathrm{lens}$ anomaly.

%% file: Results.tex
\section{Cosmological constraints on the scattering rates}\label{sec:res}

In this section we use the numerical framework described in Sec.~\ref{sec:theory} to constrain the different scattering rates involved in multi-interacting DM models. To do so, we will run MCMC scans using the parameter extraction code \textsc{MontePython}~\cite{Audren:2012wb,Brinckmann:2018cvx}. All of our parameter scans will also allow the \lcdm~parameters to vary freely, meaning the set of cosmological parameters we scan over is
\begin{equation}
\left\lbrace \omega_b, \omega_\DM, h, A_s, n_s, \tau_\mathrm{reio} \right\rbrace \ + \ \left\lbrace \text{DM model params} \right\rbrace\,,
\label{eq:params_gen}
\end{equation}
and we assume a flat prior on all \lcdm~parameters. 

First, in Sec.~\ref{sec:res_single}, we focus on the cases of only one interaction channel being activated. This will allow us to compare our results to those in the literature. In Sec.~\ref{sec:res_dual} we instead focus on all possible dual and triple interacting models, i.e. activating two or three different scattering channels. This will allow us to test if the effects of the interactions are indeed additive, as anticipated in Sec.~\ref{sec:th_imp}, or whether multiple interactions open new parameter degeneracies allowing to relax the bounds of Sec.~\ref{sec:res_single}.

In all cases, we will use the Planck 2018 baseline dataset~\cite{Aghanim:2018eyx} including temperature, polarisation and CMB lensing.\footnote{This corresponds to the high-$\ell$ TTTEEE, low-$\ell$ TT, low-$\ell$ EE, and lensing likelihoods.} Additionally, we include BAO data, using measurements of $D_V/r_{\rm drag}$ by 6dFGS at $z = 0.106$~\cite{Beutler:2011hx}, by SDSS from the MGS galaxy sample at $z = 0.15$~\cite{Ross:2014qpa}, and additionally by BOSS from the CMASS and LOWZ galaxy samples of SDSS-III DR12 at $z = 0.2 - 0.75$~\cite{Alam:2016hwk} . In addition to these BAO data sets already included in Ref. \cite{Aghanim:2018eyx}, we added new data from the DR14 eBOSS release, namely QSO clustering at $z=1.52$ \cite{Ata:2017dya}, BAO measurements from Lyman-$\alpha$ forest autocorrelation at $z=2.34$ \cite{Agathe:2019vsu}, and from cross correlation of Lyman-$\alpha$ and QSO \cite{Blomqvist:2019rah} at $z=2.35$. We refer to these datasets henceforth simply as BAO. 

\subsection{Single interaction models}\label{sec:res_single}
\begin{figure}[t!]
	\centering
	\begin{subfigure}[b]{0.298\textwidth}
		\centering
		\includegraphics[width=\textwidth]{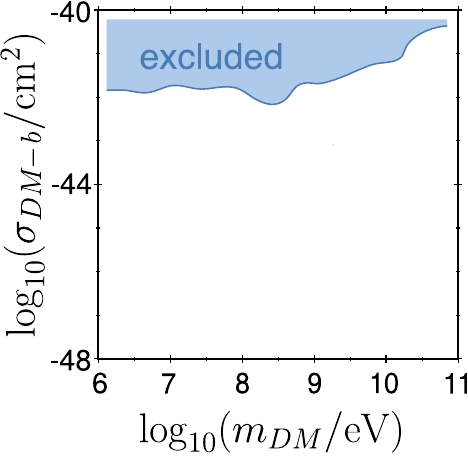}
	\end{subfigure}
	\hspace*{5mm}\begin{subfigure}[b]{0.292\textwidth}
		\centering
		\includegraphics[width=\textwidth]{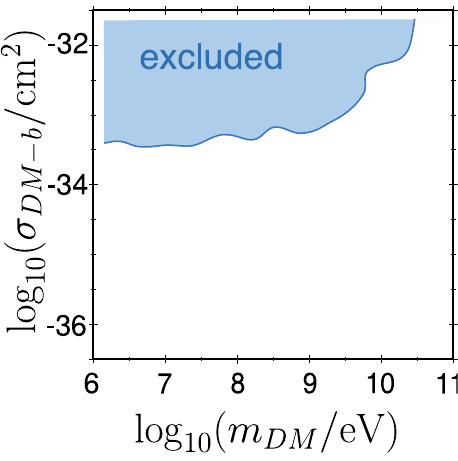}
	\end{subfigure}
	\hspace*{5mm}\begin{subfigure}[b]{0.3\textwidth}
		\centering
		\includegraphics[width=\textwidth]{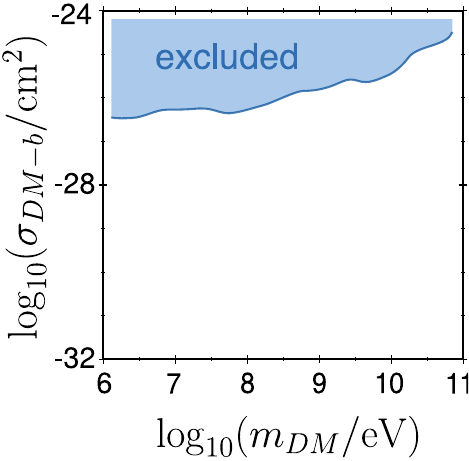}
	\end{subfigure}
	\caption{The 95\% CL exclusion limits on the DM--baryon cross section and DM mass, when assuming logarithmic priors on both parameters (see the plot range), and for three different values of the temperature scaling index $n_b\,$. \textbf{Left:} $n_b=-4$. \textbf{Middle:} $n_b=-2$. \textbf{Right:} $n_b=0$. 
	}
	\label{fig:single_b_interaction}
\end{figure}

\noindent {\bf Case of DM--baryon interactions.} We first perform three runs (corresponding to $n_b = -4, -2, 0$) in which we allow both the DM mass $m_\DM$ and the DM--baryon cross section $\sigma_{\DM \text{--}b}$ (from equation~(\ref{eq:coldmb})) to vary freely with a logarithmic prior, meaning that we have:
\begin{equation}
\left\lbrace \text{DM model params} \right\rbrace = \left\lbrace \log_{10}{m_\DM},\log_{10}{\sigma_{\DM \text{--}b}} \right\rbrace \,.
\label{eq:params_dmbmass}
\end{equation}
The corresponding results are shown in figure~\ref{fig:single_b_interaction}, where we can see that the DM mass is largely unconstrained, while there is an upper bound on the cross section. The value of the mass $m_\DM$ controls several effects in the evolution of the DM temperature and of the momentum exchange rates $\Gamma_{\DM\text{--}b}$ and $\Gamma_{b\text{--}\DM}$ (which are related to each other by a factor $\rho_\DM/\rho_b=\omega_\DM/\omega_b$). However, for $m_\DM\gg m_b$, one can always infer from equation~(\ref{eq:coldmb}) that these rates scale like $\sigma_{\DM\text{--}b}/m_\DM$, such that the bounds on $\sigma_{\DM\text{--}b}$ scale like $m_\DM$. In the opposite limit for $m_\DM \ll m_b$, equation~(\ref{eq:coldmb}) shows that the rates depend more weakly on $m_\DM$. This trend is consistent with previous results from Ref.~\cite{Xu:2018efh}, where the parameter space was scanned for three fixed values of the DM mass. Here, by considering $m_\DM$ as a free parameter in the range $1\,\mathrm{MeV}<m_\DM<100\,\mathrm{GeV}$, we get an explicit confirmation of this behaviour from the contours of figure~\ref{fig:single_b_interaction}.

Having seen that the DM mass is unconstrained in these cases, we instead choose from here on to focus on the case of $m_\DM=1\,$GeV. Additionally, given the ambiguity of upper bounds derived from a logarithmic prior with a somewhat arbitrary lower prior edge, we instead choose to focus on a flat prior on $\sigma_{\DM \text{--}b}$, leading to
\begin{equation}
\left\lbrace \text{DM model params} \right\rbrace = \left\lbrace \sigma_{\DM \text{--}b} \right\rbrace \,.
\label{eq:params_dmb2}
\end{equation}
The resulting $2\sigma$ upper bounds (95.4\,\% CL) for the cross section for this mass are shown in the first three rows of table~\ref{tab:bounds}. Given the previous discussion, these bounds can be rescaled as approximately $\sigma_{\DM\text{--}b} (1\,\mathrm{GeV}/m_\DM)$ in the limit $m_\DM\gg m_b$. 

These results can now be compared to those found in previous works. Compared to Ref.~\cite{Xu:2018efh}, our bound on the $n_b =0$ case improves by a factor $\sim1.5$, while our bound on the $n_b =-2$ case improves by a factor $\sim2$. This can be attributed to the improvement obtained when using Planck 18 instead of Planck 15, and a more complete set of BAO data. 
On the other hand, our bound on the $n_b = -4$ case degrades by a factor $\sim 1.5$ when compared to the bound in Ref.~\cite{Xu:2018efh}. This probably relates to numerical details in our exact treatment of the DM temperature evolution. Indeed, when plotting the evolution of the DM temperature like in their Figure~1, we notice small differences (only in the case $n_b=-4$) which are likely to explain our slightly looser bound. 

\begin{figure}[t!]
	\centering
	\includegraphics[width=0.32\textwidth]{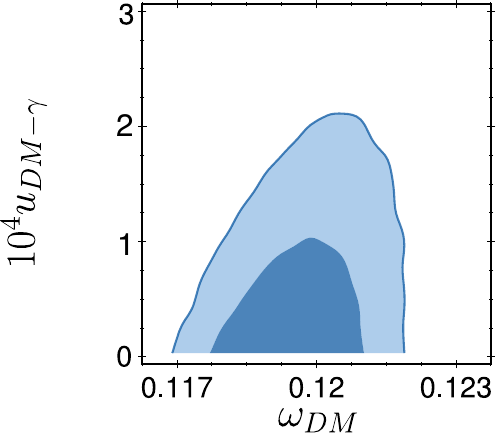}
	\caption{The posterior of the DM--photon cross section with $\omega_\DM$. The 95.4\,\% CL upper bound is $u_{\DM \gamma} < 1.8 \cdot 10^{-4}$. }
	\label{fig:single_g_interaction}
\end{figure}

\noindent {\bf Case of DM--photon interactions.}  As discussed in Sec.~\ref{sec:th_dmg}, here we only have one additional parameter $u_{\DM\text{--}\gamma}$\,, which denotes the cross section relative to the Thompson cross section and divided by $m_\DM$\,, as defined in equation~\eqref{eq:dmg}. For a fixed $u_{\DM\text{--}\gamma}$\,, varying $m_\DM$ could, in principle, have a small effect through the DM sound speed, but as discussed in Sec.~\ref{sec:th_idm} this effect is negligible on the cosmological scales probed by our datasets. As such, our set of DM model parameters to be varied together with the \lcdm~ones just consists of
\begin{equation}
\left\lbrace \text{DM model params} \right\rbrace = \left\lbrace u_{\DM\text{--}\gamma} \right\rbrace \,.
\label{eq:params_dmb}
\end{equation}
The resulting $2\sigma$ upper bound is shown in the fourth row of table~\ref{tab:bounds}, and in figure~\ref{fig:single_g_interaction}. It is looser than the result obtained in Ref.~\cite{Stadler:2018jin} based on Planck 15 (TTTEEE + lowTEB + lensing) data by about 20\%. This shift is likely due to the difference in the inferred optical depth of reionization between these datasets.

\begin{figure}[t]
	\centering
	\begin{subfigure}[b]{0.32\textwidth}
		\centering
		\includegraphics[width=\textwidth]{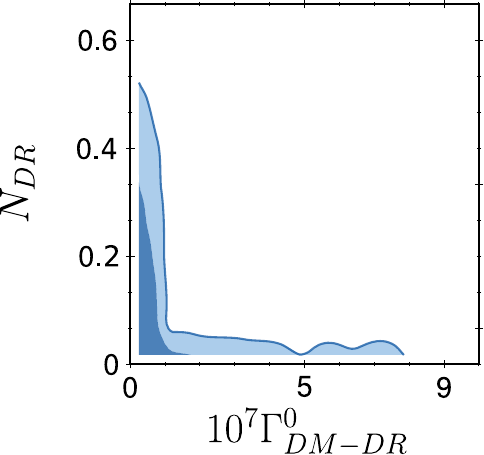}
	\end{subfigure}
	\hspace*{5mm}\begin{subfigure}[b]{0.32\textwidth}
		\centering
		\includegraphics[width=\textwidth]{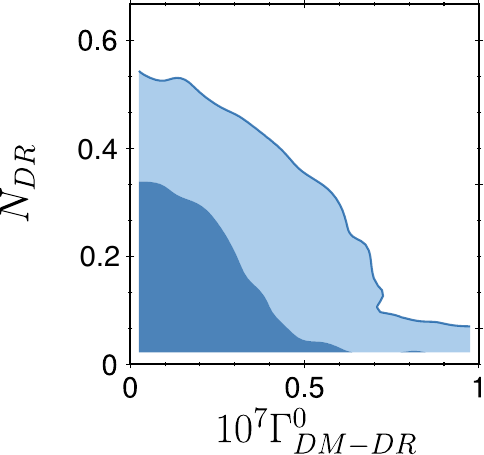}
	\end{subfigure}
	\caption{The 2D posteriors of the DM--DR momentum exchange rate and of the amount of DR characterised through $N_\DR$. \textbf{Left:} Assuming flat priors on both parameters. \textbf{Right:} Assuming an upper bound on $\Gamma^0_{\DM\text{--}\DR}$ of the form $\Gamma^0_{\DM\text{--}\DR} < 10^{-7}$.
	}
	\label{fig:single_dr_interaction}
\end{figure}

\noindent {\bf Case of DM--DR interactions.}  As discussed in Sec.~\ref{sec:th_dmdr}, we focus on the case of $n_\DR=0$ ($\Gamma_{\DM\text{--}\DR}\propto \H \propto (1+z)$ during radiation domination), and treat the DR as a perfect fluid. In this case, the value of the DM mass is irrelevant as long as $m_\DM>{\cal O}(1\,\mathrm{MeV})$,  since this implies a negligible DM sound speed. Our free parameters for the DR density and the DM--DR momentum exchange rate are
\begin{equation}
\left\lbrace \text{DM model params} \right\rbrace = \left\lbrace N_\DR, \Gamma^0_{\DM\text{--}\DR}  \right\rbrace \,.
\label{eq:params_dmdr}
\end{equation}
We first perform a run with flat priors on both parameters. The results for this run are shown in the left panel of figure~\ref{fig:single_dr_interaction}, where we can see that there is an almost bi-modal distribution, corresponding to the cases of a low interaction rate with a large amount of DR, or a high interaction rate with an almost negligible amount of DR. This can be explained by the fact that the dragging effects between DM and DR depend on both the number density of DR particles and on the DM--DR scattering rate. Thus, in the small $N_\DR$ limit, the scattering rate is practically unconstrained. When $N_\DR$ becomes sizable (typically, bigger than $\sim 0.07$), the dragging effect affects the evolution of the perturbations $(\delta_\DR, \delta_\DM)$, and the rate is bounded by the shape of the CMB spectra (and potentially also of the matter power spectrum when large scale structure data are included \cite{Lesgourgues:2015wza, Buen-Abad:2017gxg, Archidiacono:2019wdp}). At the same time, the CMB data is sensitive to $N_\DR$, which plays a role comparable to an enhanced neutrino density $\Delta N_\mathrm{eff}$. Bounds on  $N_\DR$ are, however, expected to be slightly looser than on $\Delta N_\mathrm{eff}$ for two reasons. First, in the model considered here, DR does not free-stream due to its self-interactions, and thus, has less impact on the CMB, and in particular on the acoustic peak scale \cite{Audren:2014lsa}. Second, it gets its perturbations enhanced due to the drag effect of DM, and thus, through gravitational interactions, it may push photons to cluster a bit more on small scale, counteracting the enhanced Silk damping effect induced by extra radiation. This potentially leads to a positive correlation between $N_\DR$  and $\Gamma^0_{\DM\text{--}\DR}$ that was observed in Refs.~\cite{Buen-Abad:2017gxg,Archidiacono:2019wdp} using Planck 2015 data.

We wish to avoid this bi-modality for two reasons. First, at the practical level, the reconstruction of the posterior by MCMC algorithms is difficult both in the case of bi-modality and in that of unbounded posterior distribution tails. Here both of these issues are present.
Second, as discussed in e.g. Ref. \cite{Buen-Abad:2017gxg}, it is possible to build models with $N_\DR$ much smaller than one, but under more contrived assumptions than the more generic outcome $N_\DR \sim {\cal O}(0.1-1)$. 

Thus we wish to impose a prior that will remove the ``small $N_\DR$ -- large $\Gamma^0_{\DM\text{--}\DR}$'' branch of the bi-modal posterior of figure~\ref{fig:single_dr_interaction}. There are essentially two ways to achieve this: imposing either an upper prior boundary on $\Gamma^0_{\DM\text{--}\DR}$, or a lower prior boundary on $N_\DR$.  Refs.~\cite{Lesgourgues:2015wza,Buen-Abad:2017gxg, Archidiacono:2019wdp} adopted the second strategy and imposed $N_\DR > 0.07$. The inconvenience of this choice is that the $\Lambda$CDM case is no longer recovered as a sub-case of the extended model, since this prior excludes the point $(N_\DR, \Gamma^0_{\DM\text{--}\DR} ) = (0,0)$. This may obscure the interpretation of the result. Thus we choose instead to take an upper bound on $\Gamma^0_{\DM\text{--}\DR}$ of the form $\Gamma^0_{\DM\text{--}\DR} < 10^{-7}$, which effectively cuts out the unwanted posterior tail.

Our results with such a prior are displayed in the right panel of figure~\ref{fig:single_dr_interaction} and in row five of table~\ref{tab:bounds}. Our results are relatively close to the most recent bounds on this model taken from Ref.~\cite{Archidiacono:2019wdp}, in spite of the different choice of prior and of the updated CMB and BAO data set. However, the positive correlation between $N_\DR$  and $\Gamma^0_{\DM\text{--}\DR}$ does not appear anymore: in presence of non-zero DM--DR interactions, the bounds on the DR abundance can only get stronger. We performed several intermediate runs to prove that this qualitative change with respect to the results of \cite{Buen-Abad:2017gxg, Archidiacono:2019wdp} is driven by the use of Planck 2018 data instead of Planck 2015. We conclude that the more accurate measurement of the high-$\ell$ CMB polarisation spectrum allows to better discriminate between the Silk damping effect induced by a higher $N_\DR$ and the gravitational boost effect induced by a higher $\Gamma^0_{\DM\text{--}\DR}$, and thus, by more clustered DR.
One should note that the upper bound on $\Gamma^0_{\DM\text{--}\DR}<6.2\times 10^{-8}$ at the 95.4\,\% CL reported in the table is driven mainly by the mode of \enquote{large $N_\DR$ -- small $\Gamma^0_{\DM\text{--}\DR}$}, and only very weakly depends on the choice of upper prior edge.\footnote{Even more importantly, the lack of significant change in the upper limit when adding multiple interactions is entirely driven by the data, as also visible e.g., in figure \ref{fig:triple_vs_dual_interactions}.}

\begin{table}[t]
	\centering
	\hspace*{-6mm}
	\begin{tabular}{| l | c | c | c | c | c |}
		\hline
		 Case & DM--b & DM--b & DM--b & DM--$\gamma$ & DM--DR \\
		 Index & $n_b=-4$& $n_b=-2$& $n_b=0$ & - & $n_\DR=0$ \\
		 Parameter & $\sigma_{\DM\text{--}b}$ & $\sigma_{\DM\text{--}b}$ & $\sigma_{\DM\text{--}b}$& $u_{\DM\text{--}\gamma} $ & $\Gamma^{0}_{\DM\text{--}\DR} $  \\ 
		  Units & $[10^{-41} \mathrm{cm^2}]$ & $[10^{-33} \mathrm{cm^2}]$ & $[10^{-25}\mathrm{cm^2}]$& $[10^{-4}]$ & $[10^{-8}]$  \\ \hline \hline
		DM--b ($n_b=-4$)& 2.7 & - & - & -  & -\\
		DM--b ($n_b=-2$)& - & 3.6 & - & -  & -\\
		DM--b ($n_b=~~\kern 0.125em 0$)& - & - & 2.2 & -  & -\\ \hline
		DM--$\gamma$ & - & - & - & 1.8  & -\\ \hline
		DM--DR & - & - & - & -  & 6.2 \\ \hline
		DM--b ($n_b=-4$)+DM--$\gamma$ & 2.7 & - & - & 1.9 & -\\
		DM--b ($n_b=-2$)+DM--$\gamma$ & - & 3.7 & - & 1.8  & -\\
		DM--b ($n_b=~~\kern 0.125em 0$)+DM--$\gamma$ & - & - & 2.3 & 1.7  & -\\ \hline 
		DM--b ($n_b=-4$)+DM--DR & 2.4 & - & - & - & 5.6\\
		DM--b ($n_b=-2$)+DM--DR & - & 3.1 & - & - & 6.0\\
		DM--b ($n_b=~~\kern 0.125em 0$)+DM--DR & - & - & 1.9 & -  & 6.7\\ \hline 
		DM--$\gamma$ + DM--DR & - & - & - & 1.6 & 5.5\\ \hline 
		DM--b ($n_b=-4$)+DM--$\gamma$+DM--DR & 2.5 & - & - & 1.7  & 5.4\\ 
		DM--b ($n_b=-2$)+DM--$\gamma$+DM--DR & - & 3.4 & - & 1.7  & 6.0\\
		DM--b ($n_b=~~\kern 0.125em 0$)+DM--$\gamma$+DM--DR & - & - & 1.9 & 1.5 & 6.1\\ \hline 
	\end{tabular}
	\caption{Summary of the $2\sigma$ upper bounds ($95.4\,\%$ CL) on the different interaction parameters for all of the DM interaction models considered here, assuming a mass of $m_\DM=1\,\mathrm{GeV}$. 
	}
	\label{tab:bounds}
\end{table}

\subsection{Models with multiple interactions}\label{sec:res_dual}

Since our code allows several interactions to be switched on simultaneously, we can address for the first time the question of possible degeneracies between the different interaction channels. In principle, effects from individual interactions could cancel each other, open degeneracy directions in parameter space, and allow to relax some of the bounds. Thus, to some extent, we are probing here the model dependence of CMB bounds on DM interactions.

\begin{figure}[t]
	\centering
	\includegraphics[width=0.4\textwidth]{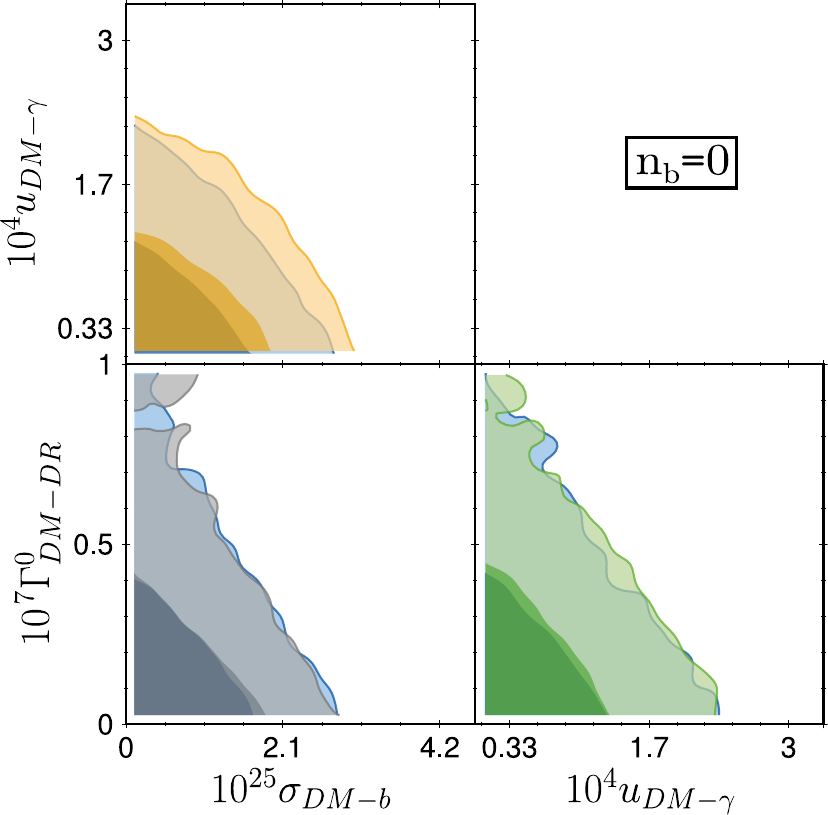}
	\includegraphics[width=0.4\textwidth]{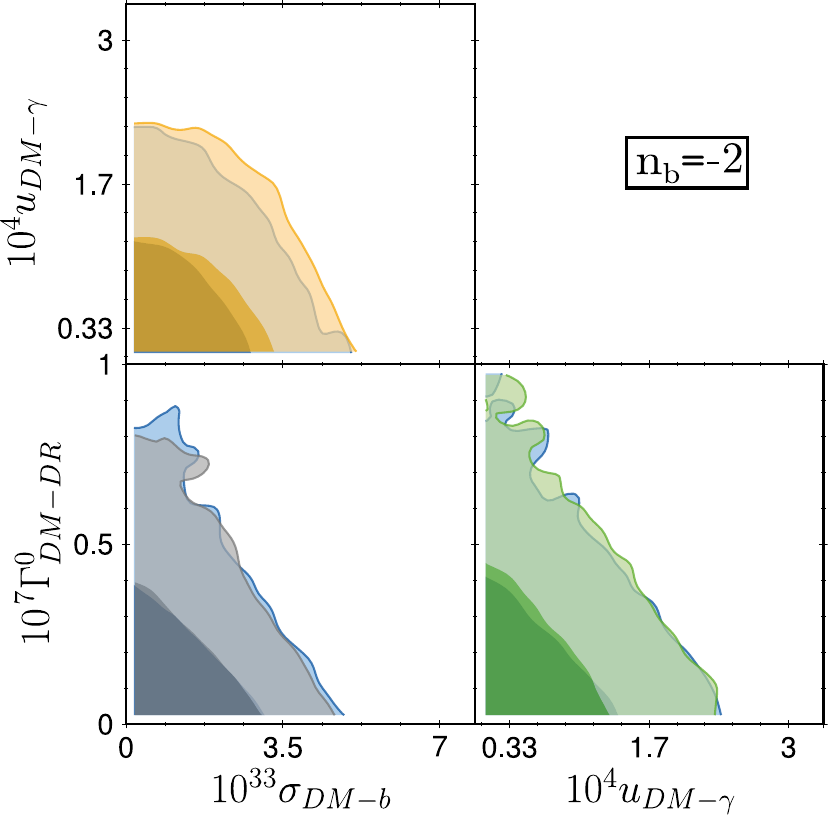}\\[2em]
	\includegraphics[width=0.4\textwidth]{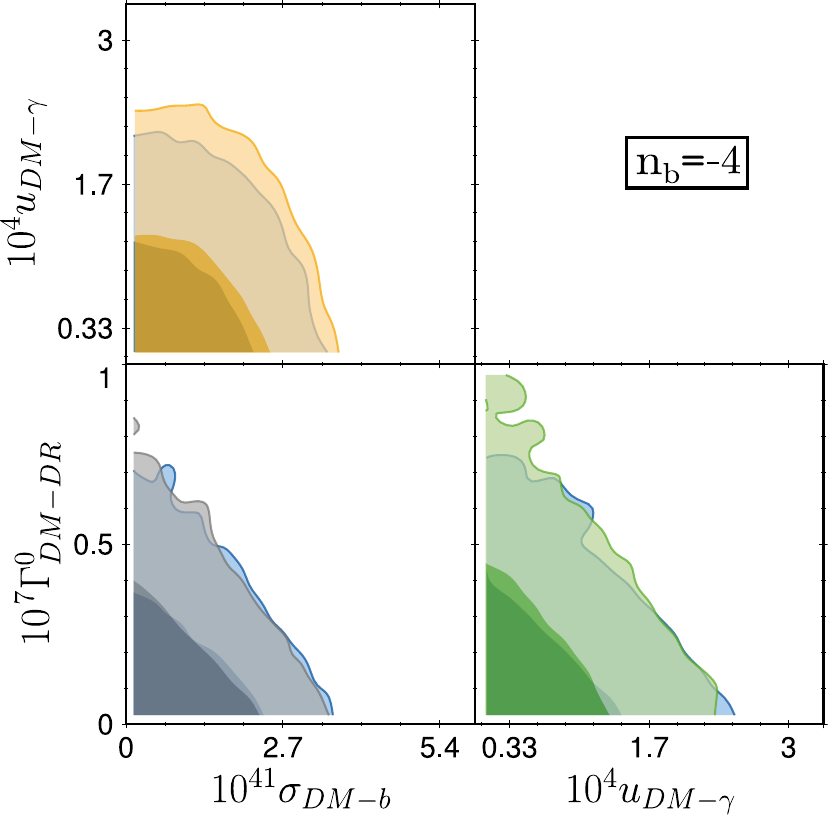}
	\hspace*{0.05\textwidth}
	\includegraphics[width=0.3\textwidth]{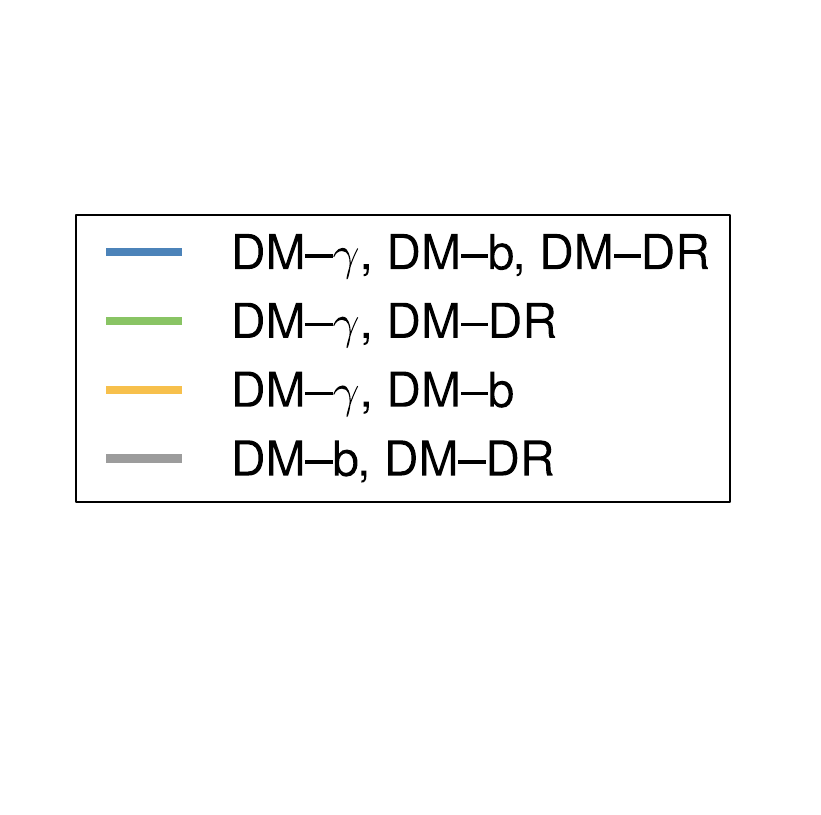}
	\caption{68.3\,\% CL and 95.4\,\% CL contours of the momentum exchange rate parameters for the various interactions. \textbf{Top Left:} Various interactions with $n_b=0$. \textbf{Top right:} Various interactions with $n_b=-2$, \textbf{Bottom:} Various interactions with $n_b=-4$.
	}
	\label{fig:triple_vs_dual_interactions}
\end{figure}

In figure~\ref{fig:triple_vs_dual_interactions} we show our results for all possible combinations of DM interacting with baryons and/or with photons (assuming a temperature-independent cross section) and/or with DR (assuming $n_\DR=0$). Each panel shows a different scaling of the DM--baryon momentum transfer cross section ($n_b=\{-4,-2,0\}$). All of the resulting $2\sigma$ upper bounds are also shown in table~\ref{tab:bounds}, which allows for quick comparison of the bounds in the different interacting scenarios.

In each panel of figure~\ref{fig:triple_vs_dual_interactions}, in the foreground we show the joint 2D confidence contours on each pair of momentum exchange rate parameters when two interactions are turned on (dual interaction model): DM--photons plus DM--DR in green, DM--photons plus DM--baryons in yellow, and DM--DR plus DM--baryons in grey. 
The contour shapes immediately convey a clear message: if there were some degeneracies, some contours would be elongated and tilted, allowing simultaneously  for two high rates compared to individual bounds. Instead, the contours are shaped like triangles or quarters-of-an-ellipse, suggesting that a larger interaction of one type typically requires a smaller interaction of the other type. This in turn implies that the various effects are additive and only their sum is constrained. Then none of the individual bounds (which correspond to the edge of the contours when one of the two parameters is zero) can be relaxed by the combined effects.

Finally, in each respective panel of figure~\ref{fig:triple_vs_dual_interactions} we show in the background in blue the 2D confidence contours on each pair of momentum exchange rate parameters when the three types of interactions are switched on simultaneously (triple interaction model). Thanks to the transparency of the contours, we see that these results are identical to those of dual interaction models. Thus there are no parameter degeneracies that only appear when the three types of effects are combined with each other.

In all of these runs, when the DM--DR interaction is turned on, we assume the same upper prior boundary $\Gamma^0_{\DM\text{--}\DR} < 10^{-7}$ as in section~\ref{sec:res_single}. We also performed additional runs without this prior, to check that even for $\Gamma^0_{\DM\text{--}\DR} > 10^{-7}$ there is no parameter degeneracy between different rates. 

In this section, we ruled out degeneracies between the parameters describing DM scattering, but we did not study possible degeneracies between these parameters and other extensions of the minimal $\Lambda$CDM model, for instance with neutrino masses larger than in the minimal hierarchy scenario considered here. The authors of~\cite{Li:2018zdm} (resp. \cite{Stadler:2018dsa}) showed that there is no degeneracy at least between the DM--baryon (resp. DM--photon) momentum exchange rate and the summed neutrino mass $\sum m_\nu$\,. Studies of degeneracies with non-standard cosmologies will be left for future work.

%% file: Tension.tex
\section{Multi-interacting dark matter and the cosmological tensions}\label{sec:tensions}

\begin{table}
	\centering
	\hspace*{-3em}
	\begin{tabular}{| l | c | c |}
		\hline
		Case & $H_0$ & $S_8$  \\ \hline
		$\Lambda${CDM}&  $67.70 \pm 0.43$& $0.825 \pm 0.011$  \\ \hline 
		DM--b ($n_b=-4$)& $67.68 \pm 0.43$ & $0.824 \pm 0.011$ \\
		DM--b ($n_b=-2$)& $67.68 \pm 0.43$ & $0.821 \pm 0.011$ \\
		DM--b ($n_b=~~\kern 0.125em 0$) & $67.70 \pm 0.43$ & $0.813 \pm 0.014$ \\ \hline
		DM--$\gamma$    & $67.70 \pm 0.43$ & $0.803 \pm 0.021$ \\ \hline
		DM--DR          & $68.73 \pm 0.96$ & $0.813 \pm 0.014$ \\ \hline
		DM--b ($n_b=-4$)+DM--$\gamma$ &  $67.68 \pm 0.43$ & $0.801 \pm 0.020$ \\
		DM--b ($n_b=-2$)+DM--$\gamma$ &  $67.69 \pm 0.44$ & $0.800 \pm 0.020$ \\
		DM--b ($n_b=~~\kern 0.125em 0$)+DM--$\gamma$  &  $67.70 \pm 0.44$ & $0.793 \pm 0.021$ \\ \hline 
		DM--b ($n_b=-4$)+DM--DR       &  $68.72 \pm 0.94$ & $0.819 \pm 0.012$ \\
		DM--b ($n_b=-2$)+DM--DR       &  $68.67 \pm 1.00$ & $0.816 \pm 0.013$ \\
		DM--b ($n_b=~~\kern 0.125em 0$)+DM--DR        &  $68.66 \pm 0.93$ & $0.810 \pm 0.014$ \\ \hline 
		DM--$\gamma$+DM--DR          &  $68.75 \pm 0.94$ & $0.799 \pm 0.020$ \\ \hline 
		DM--b ($n_b=-4$)+DM--$\gamma$+DM--DR &  $68.71 \pm 0.95$ & $0.798 \pm 0.020$ \\ 
		DM--b ($n_b=-2$)+DM--$\gamma$+DM--DR &  $68.65 \pm 0.92$ & $0.796 \pm 0.019$ \\
		DM--b ($n_b=~~\kern 0.125em 0$)+DM--$\gamma$+DM--DR  &  $68.62 \pm 0.90$ & $0.791 \pm 0.019$ \\ \hline 
	\end{tabular}
	\caption{Summary of the mean and $1\sigma$ ($68\,\%$ CL) bounds on $H_0$ and $S_8$ for all of the DM interaction models considered here. \label{tab:tensions}}
\end{table}

For each of the models studied in Sec.~\ref{sec:res}, and for the same datasets (Planck 2018 + BAO), we show in table~\ref{tab:tensions} the marginalised confidence intervals for the Hubble parameter $H_0$ and the clustering amplitude parameter $S_8$. \footnote{Note that in our MCMCs we do not include any likelihoods of weak lensing or local measurements of the Hubble parameter, as combining incompatible datasets could lead to misleading results. Instead we determine the preferred parameter regions for early and late time measurements separately and quantify the approximate tension between them.}

The first row shows the $\Lambda$CDM results for reference, using the same pipeline and datasets. We can check that the preferred range for $H_0$ is  in 4.3$\sigma$ tension with the late-time measurement of Ref.~\cite{Riess:2019cxk}, $H_0=74.03 \pm 1.42$~km/s/Mpc, while the preferred range for $S_8$ is in 2.3$\sigma$ tension with the conservative results of Ref.~\cite{Joudaki:2019pmv}, $S_8=0.762\pm0.025$.
\newpage

\begin{figure}[t!]
\centering
\includegraphics[width=0.45\textwidth]{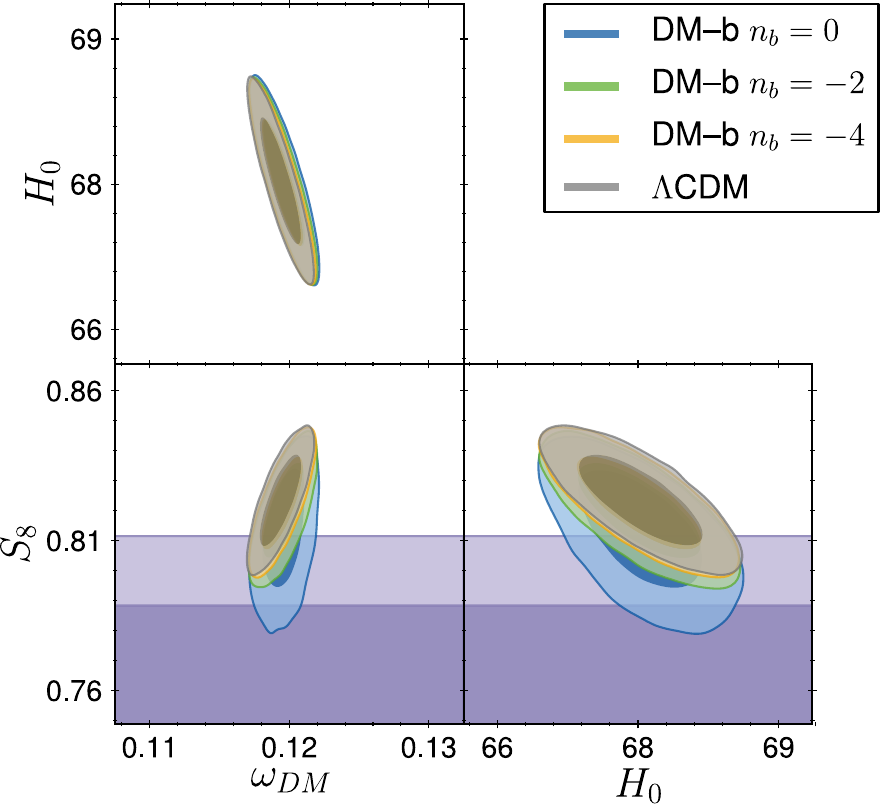}\vspace*{5mm}
\newline
\includegraphics[width=0.4\textwidth]{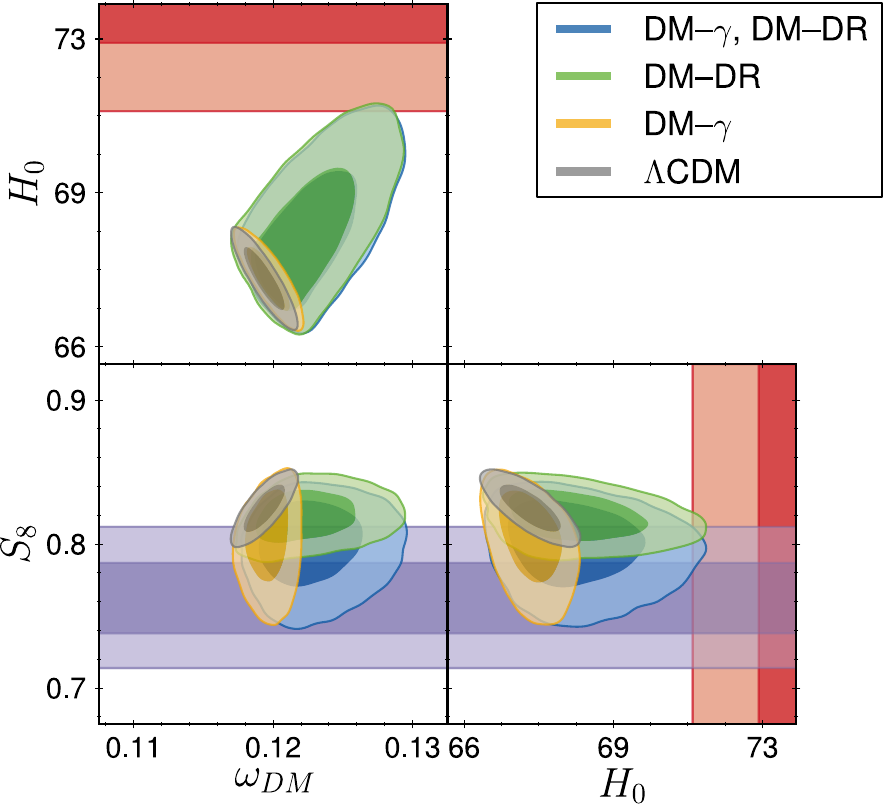}
\includegraphics[width=0.45\textwidth]{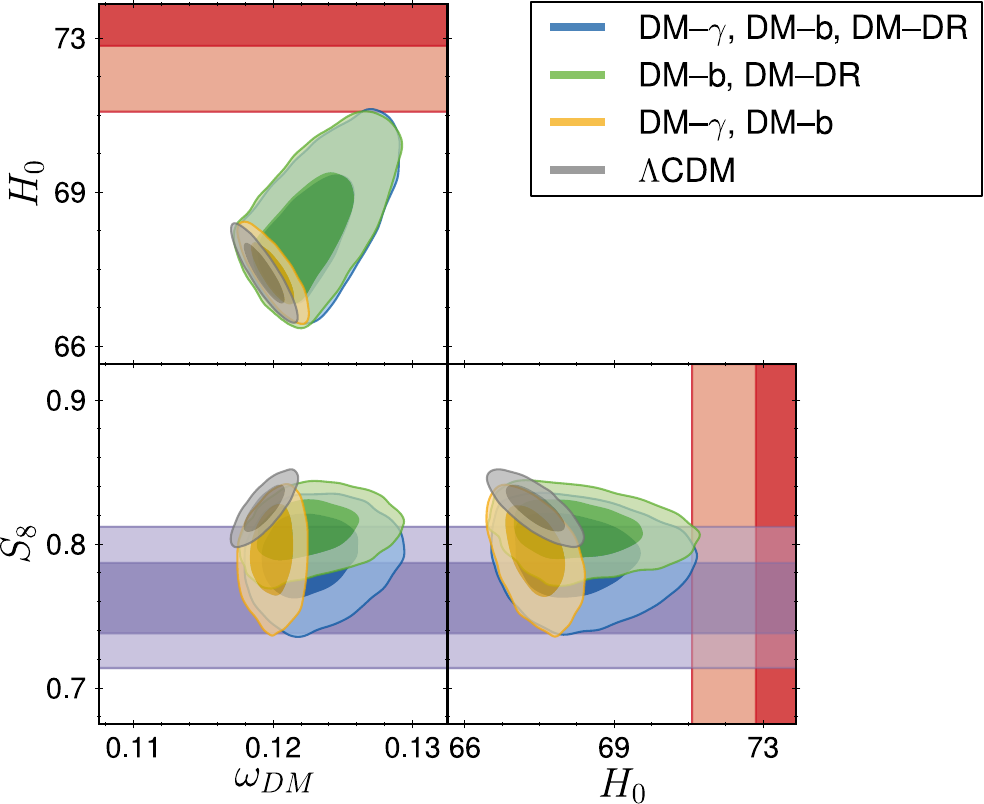}
\caption{68.3\,\% CL and 95.4\,\% CL contours of $(H_0, S_8, \omega_\DM)$,  assuming various interactions with different temperature dependencies. We show for comparison the case of the $\Lambda$CDM model, as well as the $S_8$ measurement of~\cite{Joudaki:2019pmv} in purple and the $H_0$ determination of~\cite{Riess:2019cxk} in red. \textbf{Top:} Single interactions of baryons for $n_b=\{-4,-2,0\}$. \textbf{Bottom Left:} Single interactions with DR and photons, as well as the corresponding double interaction. \textbf{Bottom Right:} Double interactions with baryons and photons or DR, as well as the triple interaction case.}
\label{fig:tensions}
\end{figure}
The subsequent rows of table~\ref{tab:tensions} show the $(H_0\,, S_8)$ predictions using Planck18 + BAO data in the case of DM--baryon interactions. The corresponding contour plots in the space $(H_0, S_8, \omega_\DM)$ are shown in figure~\ref{fig:tensions} (upper panel). Predictions for $H_0$ are unaffected by this type of interaction, which does not incorporate any mechanism to counteract an increase in $H_0$. The value of $S_8$ is significantly affected only in the $n_b=0$ case, that is, when the DM--baryon cross-section quickly decreases with time, and is thus potentially very large in the early universe. In this case, CMB bounds are compatible with values of the momentum exchange rate that lead to a suppression of the matter power spectrum on scales that are relevant for $S_8$. Note that the inclusion of Lyman-$\alpha$ data would result in stronger bounds on the momentum exchange rate \cite{Xu:2018efh}, which would restrict the possibility to lower $S_8$. 

\newpage
The next line in table~\ref{tab:tensions} shows that DM--photon interactions can efficiently reduce $S_8$. In this case, the matter power spectrum is suppressed on small scales because DM density fluctuations remain as small as photon fluctuations as long as the two species are coupled (there are even acoustic oscillations in the coupled DM--photon fluid). As already discussed in Ref.~\cite{Stadler:2018jin}, the CMB puts bounds on $u_{\DM\text{--}\gamma}$ that are compatible with a reduction of the matter power spectrum on scales relevant for $S_8$ (see figure 5 in~\cite{Stadler:2018jin}). We find that this is still the case with our Planck 18 + BAO dataset: the $S_8$ tension gets reduced from the 2.3$\sigma$ to the 1.3$\sigma$ level by the DM--photon interaction. We should, however, keep in mind that our comment on the DM--baryon case applies also to this case: the reduction of $S_8$ might become marginal if we used Lyman-$\alpha$ data to put stronger bounds on $u_{\DM\text{--}\gamma}$\,.

The next line in table~\ref{tab:tensions} confirms the findings of Refs.~\cite{Lesgourgues:2015wza, Buen-Abad:2017gxg, Archidiacono:2019wdp}  in that the DM-DR interaction model with $n_\DR=0$ can reduce both tensions by a moderate amount (from 4.3$\sigma$ to 3.1$\sigma$ for $H_0$, and from 2.3$\sigma$ to 1.8$\sigma$ for $S_8$). The increase in $H_0$ is mainly due to the presence of self-interacting DR, and the decrease in $S_8$ is due to the drag effect of DR on DM. Note that the authors of \cite{Lesgourgues:2015wza, Buen-Abad:2017gxg, Archidiacono:2019wdp} found that both DR self-interactions and DR--DM interactions help reaching higher values of the total radiation density than in a plain $\Lambda$CDM$+N_\mathrm{eff}$ model with additional free-streaming degrees of freedom -- and thus, also, higher values of $H_0$. As reported in section 3.1, with Planck 2015 replaced by Planck2018 data, the role of the DM--DR interactions is no longer obvious for this mechanism to work. DM--DR interactions still play a role in the reduction of $S_8$.

Since DM--photon interactions offer the most efficient way to reduce $S_8$, and DM--DR to increase $H_0$, we should check the predictions of the combined model for the cosmological tensions. The results are shown in the line labelled \enquote{DM--$\gamma$+DM--DR} in table~\ref{tab:tensions}, and are well summarised by figure~\ref{fig:tensions} (bottom left panel). In this case, the tensions get simultaneously reduced from 4.3$\sigma$ to 3.1$\sigma$ for $H_0$, and from 2.3$\sigma$ to 1.2$\sigma$ for $S_8$. The figure shows very clearly that the confidence contours of the combined model incorporate a large region of parameter space with high $H_0$ and  low $S_8$ which would be incompatible with the data in each single interaction model.  For instance, the case $(H_0=71\, \mathrm{km/s/Mpc}, S_8=0.77)$ lies within the 95\,\% CL marginalised contours of the dual interaction model, but in none of the 95\,\% CL contours of the single interaction model. Of course, we should keep in mind that this is done at the expense of introducing three new parameters. 

The bottom right panel of figure~\ref{fig:tensions} finally confirms that switching on the DM--baryon interactions with $n_b = 0$ -- which was  shown to be the DM--baryon case with the largest impact on $S_8$ -- on top of the other two channels has no further impact on the cosmological tensions.

%% file: Discussion.tex
\newpage
\section{Discussion}\label{sec:conc}
The non-detection of DM by current experiments combined with a series of unexplained tensions in cosmological data, provide diverse but reasonable motivations for investigating the cosmological signatures of a non-trivial dark sector of particle physics.

In this work, we have shown that it is possible to gather multiple channels for DM elastic scattering with other species within a single Boltzmann code, with a consistent treatment of the thermal evolution and of several tight-coupling regimes. Our code features DM--baryon, DM--photon, and DM--DR interactions (this last one already present in \class v2.9~\cite{Archidiacono:2017slj, Archidiacono:2019wdp}), and allows multiple interaction channels of the DM species to be switched on simultaneously without making the Boltzmann code significantly slower. This code will constitute the version 3.1 of \class, and its public release will follow the publication of this paper.

We have investigated the cosmological effects of multiple DM interactions. For this, we focused on joint constraints on the various momentum exchange rates, and on the role of multiple interaction models in possibly alleviating the cosmological tensions. Our analysis yields two main results. The first one is best summarised by \mbox{figure~\ref{fig:triple_vs_dual_interactions}}. These plots show at the first glance that when multiple interactions are switched on, there are no counteracting effects leading to parameter degeneracies and to a relaxation of CMB bounds on individual momentum exchange rates.

Our second result is demonstrated by figure~\ref{fig:tensions} and table~\ref{tab:tensions}. We find that the combination of several interaction channels can help to reduce the cosmological tensions. In a set up where a single DM relic interacts feebly with the visible sector (through DM--photon scattering) and with dark relics (assumed to be relativistic), the CMB and BAO data are compatible with large values of $H_0$ and low values of $S_8$, such that, for instance, models with  $H_0=71\, \mathrm{km/s/Mpc}$ and $S_8=0.77$ lie within the 95\,\% CL marginalised two-dimensional contours. We acknowledge that this model requires three extra free parameters with respect to the minimal $\Lambda$CDM model, and does not completely eliminate the Hubble tension, which is still of the order of 3.1$\sigma$ with respect to the direct measurement of Ref.~\cite{Riess:2019cxk}.

The release of our code paves the way towards the study of more complicated dark sector models, in which there could be multiple DM relics, decays within the dark sector, effects of inelastic scattering, or transitions between energy levels if the dark sector contains dark atoms. Some of these models would require only minimal modifications to our code: for instance, one could easily explore a different dependence of the momentum exchange rate over the dark sector  temperature(s), or nest the DM equations within a loop in order to simulate several DM relics each with different properties. In any case, our code already provides the basic infrastructure for simulating extended dark sectors due to the generic differential equation solver. Studies of such extended dark sectors may bring more convincing explanations of the $H_0$ and $S_8$ tensions, and potentially of other unexplained observations such as the EDGES anomaly or the small scale crisis.

%% file: App_Equation.tex
\section{Main equations}
\label{App:eqs}
Here we list the main equations that have been modified to account for multi-interacting DM. For such models, the gauge transformations are straightforward: the perturbation equations are identical in different gauges up to the few terms featuring metric perturbations. Thus, we only write here the Newtonian gauge equation, although our code also works (and gives the same results) in the synchronous gauge.\footnote{The usual synchronous gauge is defined to be comoving at all times with decoupled CDM. 
When the user chooses to split DM between a decoupled CDM component and an IDM component, our synchronous gauge is defined to be comoving with the former. When all the DM is assumed to be interacting, our code adds automatically a negligible fraction of decoupled CDM and sticks to the convention $\theta_\mathrm{CDM}=0$.}
We recall that primes denote derivatives with respect to conformal time. For specifics about the \class implementation, we refer to App.~\ref{App:num}.
\subsection*{Background}
At the background level, the only relevant quantities are the energy density evolution equations of DM and DR, given by
\begin{align}
	\rho_{\DM}(a) =& \rho_\mathrm{crit} \,\Omega_{\DM,0}\, \left(a/a_0\right)^{-3}\,, \\
	\rho_{\DR}(a) =& \rho_\mathrm{crit} \,\Omega_{\DR,0}\, \left(a/a_0\right)^{-4} \,,
\end{align}
where $\Omega_{x,0}$ denotes the relic abundance of species $x$ today. These expressions are based on the massless nature of DR and the non-relativistic nature of DM, and explicitly assume that the DR interactions do not significantly inject energy. The other species behave in the same way as in the standard $\Lambda$CDM model.

\subsection*{Thermodynamics}
The presence of DM--baryon interactions affects the mean baryon temperature. The new evolution for the baryon temperature then becomes
\begin{equation}
{T}_{b}^\prime = -2\H T_{b} - \frac{2\mu_b}{m_e} \Gamma_{b\text{--}\gamma} (T_b - T_\g) - \frac{2\mu_{b}}{m_{\DM} + m_{b} } \Gamma_{b\text{--}\DM}(T_b - T_\DM) \, .
\end{equation}
Here $\Gamma_{b\text{--}\gamma}$ is the conformal baryon--photon momentum exchange rate due to Thomson scattering, such that $\Gamma_{b\text{--}\gamma}= \frac{4\rho_\g}{3\rho_{b}}
\Gamma_{\gamma\text{--}b}$, where 
\begin{equation}
\Gamma_{\gamma\text{--}b} = a \, \sigma_T n_e\,,
\end{equation}
with $n_e$ the free electron number density and $\sigma_T$ the Thomson cross section (the parameter $\Gamma_{\gamma\text{--}b}$ is called $\kappa'$ within \class). 

$\Gamma_{b\text{--}\DM}$ is the conformal baryon--DM momentum exchange rate, such that  $\Gamma_{b\text{--}\DM}=  \frac{\rho_{\DM}}{\rho_b} \Gamma_{\DM\text{--}b}$, where $\Gamma_{\DM\text{--}b}$ is given by equation \eqref{eq:coldmb} and is denoted $R_\chi$ in Ref.~\cite{Xu:2018efh}.
The mass of each species $x$ is denoted by
$m_x$ and the baryon mean molecular weight is given by $\mu_{b}$\,.

Furthermore, all of the interacting species have an impact on the DM temperature, which is not necessarily negligible and needs to be evolved together with the baryon temperature, as discussed in Sec.~\ref{sec:th_idm}. The DM temperature evolution is given by 
\begin{equation}
\begin{split}
{T}_{\DM}^\prime = -2\H T_{\DM} &- 2 \Gamma_{\DM\text{--}\gamma}(T_\DM - T_{\g})\\
& - 2 \Gamma_{\DM\text{--}\DR} (T_\DM - T_{\DR}) \\
&- \frac{2m_{\DM}}{m_{\DM} + m_{b} } \Gamma_{\DM\text{--}b} (T_\DM - T_{b})\, ,
\end{split}
\end{equation}
where $\Gamma_{\DM\text{--}\gamma}$ is the conformal DM--photon momentum exchange rate given in equation~\eqref{eq:gamma_dm_g}, such that  $\Gamma_{\DM\text{--}\gamma}=\frac{4\rho_\gamma}{3 \rho_\DM}\Gamma_{\gamma\text{--}\DM}$ ($\Gamma_{\gamma\text{--}\DM}$ is denoted $\dot{\mu}$ in Ref.~\cite{Stadler:2018jin}). $\Gamma_{\DM\text{--}\DR}$ is the conformal DM--DR momentum exchange rate from equation~\eqref{eq:coldmdr}, such that $\Gamma_{\DM\text{--}\DR}= \frac{4 \rho_\DR}{3 \rho_\DM} \Gamma_{\DR\text{--}\DM}$ (same notations as in Ref.~\cite{Archidiacono:2019wdp} up to a sign flip).\footnote{In Ref.~\cite{Cyr-Racine:2015ihg}, the rates $\dot{\kappa}_{\DM\text{--}\DR}$, $\dot{\kappa}_{\DR\text{--}\DM}$, $\dot{\kappa}_{\DR\text{--}\DR}$ were all negative, because they stand for the time derivative of visibility functions. In Ref.~\cite{Archidiacono:2019wdp}, the same negative rates were called $\Gamma_{\DM\text{--}\DR}$, $\Gamma_{\DR\text{--}\DM}$, $\Gamma_{\DR\text{--}\DR}$. Here we define all our rates to be positive, in order to adopt more homogeneous conventions across different interaction channels. Thus our notations include a sign flip with respect to Ref.~\cite{Archidiacono:2019wdp}.}
Once the DM temperature is known, the DM sound speed follows as
\begin{equation}
	c_{\DM}^2 = \frac{k_B T_{\DM}}{m_{\DM}} \left(1 - \frac{1}{3} \pdv{\ln T_{\DM}}{\ln a} \right) \, .
\end{equation}
We do not consider changes in the photon or DR temperature other than the $(1+z)$ scaling due to the adiabatic expansion, as the additional scatterings can be described as very small spectral distortions to the photon and DR phase space distributions and are assumed to be negligible (see Sec.~\ref{sec:th_dmdr}).

We summarise the correspondence between different notations for the momentum exchange rates in table~\ref{tab:notations}.
\begin{table}
\begin{center}
\begin{tabular}{|c|c|c|}
\hline
This work & \class & Other works\\
\hline
$\Gamma_{\g\text{--}b}$ & \texttt{dkappa} & $\dot{\kappa}$ in Ref.~\cite{Stadler:2018jin} \\
$\Gamma_{\DM\text{--}b}$ & \texttt{R\_idm\_b} & $R_\chi$ in Ref.~\cite{Dvorkin:2013cea}\\
$\Gamma_{\g\text{--}\DM}$ & \texttt{dmu\_idm\_g} & $\dot{\mu}$ in Ref.~\cite{Stadler:2018jin}\\
$\Gamma_{\DR\text{--}\DM}$ & \texttt{dmu\_idm\_dr} & $- \dot{\kappa}_{\DR\text{--}\DM} = -\Gamma_{\DR\text{--}\DM}$ in Refs.~\cite{Cyr-Racine:2015ihg,Archidiacono:2019wdp}\\
$\Gamma_{\DR\text{--}\DR}$ & \texttt{dmu\_idr} & $- \dot{\kappa}_{\DR\text{--}\DR} = -\Gamma_{\DR\text{--}\DR}$ in Refs.~\cite{Cyr-Racine:2015ihg,Archidiacono:2019wdp}\\
\hline
\end{tabular}
\end{center}
\caption{Correspondence between the notations of this work, of \class, and of other papers.
\label{tab:notations}}
\end{table}

\subsection*{Perturbations}
For baryons, the continuity equation is unchanged, while the Euler equation features the baryon--DM momentum exchange rate:
\begin{align}
\delta_{b}^\prime =& -\theta_{b} + 3\phi'~, \\ 
	\theta_{b}^\prime =& -\H \theta_{b} + c_{b}^2 k^2 \delta_{b} + k^2 \psi - \Gamma_{b\text{--}\g} (\theta_b - \theta_\g) - \Gamma_{b\text{--}\DM} (\theta_{b} - \theta_\DM) ~,
\end{align}
where $c_b$ is the usual baryon sound speed and $\H=\dot{a}/a$. Likewise, the photon Boltzmann equations are modified to account for the DM--photon interactions:
\begin{align}
\delta'_{\g} =& - \frac{4}{3}\theta_{\g} + 4\phi'~,\\ 
\theta'_{\g} =& \, k^2\left(\frac{1}{4}\delta_\g - \sigma_\g \right) + k^2 \psi - \Gamma_{\g\text{--}b}(\theta_\g - \theta_b ) - \Gamma_{\g\text{--}\DM}(\theta_{\g} - \theta_\DM)~, \\
\sigma'_{\g} =& \frac{4}{15}\theta_\g - \frac{3}{10}kF_{\g3} - \frac{9}{10}(\Gamma_{\g\text{--}b} + \Gamma_{\g\text{--}\DM}) \sigma_\g  + \frac{1}{20}(\Gamma_{\g\text{--}b} + \Gamma_{\g\text{--}\DM}) (G_{\g 0} + G_{\g 2})~,\\
F'_{\g \ell} =& \frac{k}{2\ell+1}\left[\ell F_{\g (\ell-1)} - (\ell+1) F_{\g (\ell+1)}\right] - (\Gamma_{\g\text{--}b} + \Gamma_{\g\text{--}\DM}) F_{\g \ell}, \quad \ell \geq 3~, \\
G'_{\g 0} =& -k G_{\g1} - \frac{1}{2}(\Gamma_{\g\text{--}b} + \Gamma_{\g\text{--}\DM})(G_{\g0} - F_{\g2} - G_{\g 2})~, \\
G'_{\g 1} =& \frac{k}{3} (G_{\g0} - 2 G_{\g 2}) - (\Gamma_{\g\text{--}b} + \Gamma_{\g\text{--}\DM})G_{\g1}~, \\
G'_{\g 2} =& \frac{k}{5}(2G_{\g 1} - 3G_{\g 3}) + \frac{\Gamma_{\g\text{--}b} + \Gamma_{\g\text{--}\DM}}{10}(F_{\g2} + G_{\g0} + G_{\g 2}) - (\Gamma_{\g\text{--}b} + \Gamma_{\g\text{--}\DM})G_{\g 2}~,\\
G'_{\g \ell} =& \frac{k}{2\ell+1}\left(\ell G_{\g(\ell-1)} - (\ell+1)G_{\g(\ell+1)}\right) - (\Gamma_{\g\text{--}b} + \Gamma_{\g\text{--}\DM})G_{\g \ell}, \quad \ell\geq 3 ~ ,
\end{align}
and the truncation formula at some $\ell_\mathrm{max}$ also contains the sum of the two rates $\Gamma_{\g\text{--}b} + \Gamma_{\g\text{--}\DM}$.

In general, the DR perturbations also obey a Boltzmann hierarchy that involves the DR--DM and DR--DR interactions:
\begin{align}
\delta'_{\DR} =& - \frac{4}{3} \theta_{\DM} + 4\phi'~, \\ 
\theta'_{\DR} =& \, k^2\left(\frac{1}{4}\delta_{\DR} - \sigma_{\DR}\right) + k^2 \psi - \Gamma_{\DR\text{--}\DM} (\theta_{\DR} - \theta_{\DM})~, \\
\Pi'_{\DR,\ell} =& \frac{k}{2\ell+1}( \ell \Pi_{\DR, \ell-1} - (\ell+1) \Pi_{\DR,\ell+1}) - (\alpha_\ell \Gamma_{\DR\text{--}\DM} + \beta_\ell \Gamma_{\DR\text{--}\DR})\Pi_{\DR,\ell} ~,
\end{align}
where the terms $\alpha_\ell$, $\beta_\ell$ are related to the DR angular coefficients, as defined in Refs.~\cite{Cyr-Racine:2015ihg,Archidiacono:2019wdp}. When the user requests strongly self-interacting DM, the hierarchy is truncated at $\ell=1$ with $\sigma_\DR=0$.
Finally, the DM perturbations feel the presence of all interactions (as briefly described in Sec.~\ref{sec:theory}), and the corresponding continuity and Euler equations are given by
\begin{align}
\delta'_{\DM} =& -\theta_{\DM} + 3\phi'~, \\ 
\theta'_{\DM} =& -\H \theta_{\DM} + c_{\DM}^2 k^2 \delta_{\DM} + k^2 \psi 
	\begin{aligned}[t]
	&- \Gamma_{\DM\text{--}\g} (\theta_{\DM} - \theta_{\g}) \\
	&- \Gamma_{\DM\text{--}b} (\theta_{\DM} - \theta_{b}) \\
	&- \Gamma_{\DM\text{--}\DR}(\theta_{\DM} - \theta_{\DR}) \, .
	\end{aligned}
\end{align}

\subsection*{Source functions}
The presence of DM--photon interactions also impacts the CMB temperature and polarisation anisotropy source functions\footnote{Note that in \class the three scalar source functions $S_T^0$, $S_T^1$, $S_T^2$ are never derived (neither analytically nor numerically) and never combined with each other for the calculation of the temperature spectrum. They are just directly convolved with three different radial functions, as suggested by the line-of-sight method. Combining the source functions together is only required when the line-of-sight formula is rearranged through integrations by part, a step not assumed by \class, as explained in \cite{Lesgourgues:2013bra}.} as there can be additional re-scattering of the photons from DM along the line of sight. Thus, they will receive additional terms coming from the interaction rate $\mu' = \Gamma_{\g\text{--}\DM}$:
\begin{align}
	\kappa =& - \int_{\tau_0}^{\tau} \kappa'd\tau , \quad\mu = -\int_{\tau_0}^{\tau} \mu'  d\tau \\ 
	g(\tau) =& (\kappa'+\mu') e^{-\kappa - \mu} \\ 
	S_T^0 =& g \left(\frac{1}{4}\delta_\g + \phi \right) + 2e^{-\kappa - \mu} \phi' \\
	& +  \frac{1}{k^2}\Big[ g(\kappa'\theta_b + \mu' \theta_\DM) + e^{-\kappa - \mu}(\kappa'' \theta_b + \mu'' \theta_\DM + \kappa'\theta_b' + \mu'\theta_\DM') \Big] \nonumber \\
	S_T^1 =& e^{-\kappa - \mu}k(\psi - \phi ) \\ 
	S_T^2 =& \frac{1}{8}g\left(G_{\gamma 0}+G_{\gamma 2} + 2 \sigma_\gamma \right) \,.
\end{align}
The corresponding formulas in the synchronous gauge are simply found \cite{Ma:1995ey} by  replacing \\ \mbox{$\phi \to \eta-\H \alpha$, $\psi \to \alpha'+\H \alpha$, and $\theta_x \to \theta_x+\alpha k^2$ for $x \in \{b,\DM\}$, where $\alpha = \frac{1}{2k^2} (h'+6\eta)$}.

\subsection*{Baryon--Photon tight-coupling approximation}
One important point of the IDM, as already described in Sec.~\ref{sec:th_idm}, is the impact these interactions have on the tight-coupling regime, which will feel both the effects of DM--photons and DM--baryon interactions. 

At first order in the tight-coupling approximation, the derivative of the photon-baryon slip $\Theta^{\text{tca}}_{\g b}\equiv \theta_\gamma-\theta_b$ is then given by

\begin{align}
	\Theta'^{\text{tca}}_{\g b} =& \left(\frac{\tau'_c}{\tau_c} - \frac{2\H}{1+R}\right)\Theta^{\text{tca}}_{\g b}   \\
	&  - \frac{\tau_c}{1+R} \Bigg[-\frac{a''}{a}\theta_b + k^2\left(-\frac{1}{2}\H \delta_g + {c'}_b^2\delta_b + c_b^2 \delta'_b + \frac{1}{4}\delta'_g + \H\psi\right) \nonumber \\
	& - \Gamma_{\DM\text{--}\g}(\theta'_\DM - \theta'_\g) - \Gamma_{\DM\text{--}b}\frac{\rho_\DM}{\rho_b} \Big( (\theta'_\DM - \theta'_b) + \Big(\H + \frac{{\Gamma'}_{\DM\text{--}b}}{\Gamma_{\DM\text{--}b}}\Big)(\theta_\DM - \theta_b)\Big) \Bigg] \,.\nonumber 
\end{align}
Here we have defined $\tau_c = 1/\kappa'$. 

\noindent This term will then affect the photon and baryon expressions in the following way:
\begin{align}
	\theta'_b =& -\frac{1}{1+R}\Bigg[\H\theta_b - c_b^2 k^2 \delta_b - k^2R\left(\frac{1}{4}\delta_\g - \sigma^{\text{tca}}_\g\right)   +  R {\Theta'^{\text{tca}}_{\g b}} \\
	& - \Gamma_{\DM\text{--}b} \frac{\rho_\DM}{\rho_b}(\theta_\DM - \theta_b)  - \Gamma_{\DM\text{--}\g} R (\theta_g - \theta_\DM) \Bigg]  + k^2\psi\nonumber \\
\theta'_g =& - \frac{1}{R} \left(\theta'_b + \H \theta_b - k^2 c_b^2 \delta_b\right) + k^2\left(\frac{1}{4}\delta_\g - \sigma^{\text{tca}}_\g \right) \\
	& - \Gamma_{\DM\text{--}\g}(\theta_g - \theta_\DM) + \frac{1}{R} \frac{\rho_\DM}{\rho_b} \Gamma_{\DM\text{--}b} (\theta_\DM - \theta_b) + \frac{R}{1+R}k^2\psi \,, \nonumber 
\end{align}
where $R = \frac{4\rho_\g}{3\rho_b}$ and the photon shear is $\sigma^{\text{tca}}_\g  = \frac{16}{45} \theta_\g \frac{\tau_c}{1+\Gamma_{\g\text{--}\DM}\tau_c}$ at first order.

\subsection*{Dark Matter--Dark Radiation tight-coupling approximation}
Finally, the DM and DR can also be tightly coupled, and this will also need to be modified to account for the baryon and photon interactions. At first order in the tight-coupling approximation, the derivative of the DM--DR slip $\Theta^{\text{tca}}_{\DM \DR}\equiv \theta_\DM-\theta_\DR$ is then given by
\begin{align}
	\Theta'^{\text{tca}}_{\DM \DR} =& \left(n - \frac{2}{1+R}\right)\H \Theta^{\text{tca}}_{\DM \DR}   \\
	&  - \frac{\tau_c}{1+R} \Bigg[-\frac{a''}{a}\theta_{\DM} + \H k^2\left(c^2_{\DM}\delta_{\DM} - \frac{1}{2}\delta_{\DR} - \psi\right) \nonumber \\
	& + k^2\left(c'^2_{\DM}\delta_{\DM} + c^2_{\DM} \delta'_{\DM} - \frac{1}{4}\delta'_{\DR}\right)  \nonumber \\
	& - (\H \Gamma_{\DM\text{--}b} + \Gamma'_{\DM\text{--}b})(\theta_\DM - \theta_b) - \Gamma_{\DM\text{--}b} (\theta'_\DM - \theta'_b) \nonumber \\
	& + 2\H \frac{4\rho_{\g}}{3\rho_\DM} \Gamma_{\DM\text{--}\g} (\theta_\DM - \theta_\g) - \frac{4\rho_{\g}}{3\rho_\DM}\Gamma_{\DM\text{--}\g}(\theta'_\DM - \theta'_\g) \Bigg] \,. \nonumber  
\end{align}
The Euler equations are
\begin{align}
\theta'_\DM =&  \frac{1}{1+R} (-\H \theta_\DM + k^2 c_{\DM}^2\delta_\DM - \Gamma_{\DM\text{--}b} (\theta_\DM - \theta_b) - \Gamma_{\DM \text{--}\g}(\theta_\DM - \theta_\g)) \nonumber \\ 
 & {} +  \frac{R}{1+R} (-k^2(\sigma_\DR - \frac{1}{4}\delta_{\DR})) + \frac{R}{1+R} \Theta'^{\text{tca}}_{\DM \DR} + k^2 \psi \,,
\end{align}
where $\theta'_\DR$ is, by definition 
\begin{equation}
\theta'_\DR =\theta'_\DM - \Theta'^{\text{tca}}_{\DM \DR} \, .
\end{equation}

%% file: App_Decoupling.tex
\newpage
\section{Decoupling redshifts}
\label{App:dec}

\subsection*{Dark matter -- baryon decoupling}

The conformal DM--baryon momentum exchange rate is given by equation~(\ref{eq:coldmb}).
For typical models, at high redshift, the term between parenthesis is dominated by 
\begin{equation}
\frac{T_b}{m_b} + \frac{T_\DM}{m_\DM} \simeq \frac{T_\gamma}{m_b} \left(\frac{m_\DM+m_b (T_\DM/T_\gamma)}{m_\DM} \right)\,,
\end{equation} 
where the ratio $T_\DM/T_\gamma$ is negligible as long as DM is decoupled form baryons, and close to one when DM is tightly coupled to baryons of temperature $T_b\simeq T_\gamma$. Then we can rewrite equation~(\ref{eq:coldmb}) as:
\begin{align}
\Gamma_{\DM\text{--}b} = & 2.87\times10^{-28} \left(2.50\times10^{-5}\right)^\frac{n_b}{2}\left(\frac{\omega_b}{0.0224}\right) (1+z)^\frac{n_b+5}{2} \nonumber \\
& c_{n_b} \frac{\left( 1+ \frac{1}{R_\DM}\frac{T_\DM}{T_\gamma} \right)^\frac{n_b+1}{2}}{\left(1+R_\DM\right)} \left(\frac{\mathcal{F}_\mathrm{He}}{0.76}\right)  \left(\frac{\sigma_{\DM-b}}{10^{4n_b-25} \mathrm{cm}^2}\right) a_0 \, \mathrm{s}^{-1}\,,
\label{eq:gamma_dm_b}
\end{align}
where $\omega_\mathrm{b}=\Omega_\mathrm{b} h^2$ is the baryon density parameter, and $R_\DM= m_\DM/m_b$.
During radiation domination the conformal Hubble rate can be expressed as a function of the effective neutrino number $N_\mathrm{eff}$ (equal to 3.044 in the standard cosmological model \cite{Froustey:2020mcq,deSalas:2016ztq}): 
\begin{equation}
\H = 2.10\times 10^{-20} \left(
\frac{1+ N_\mathrm{eff}f_{1\nu}}{1+3.044\, f_{1\nu}}
\right)^{1/2} (1+z) \, a_0 \, \mathrm{s}^{-1}\,,
\label{eq:aHrd}
\end{equation}
where we introduced the neutrino-to-photon density ratio (in the instantaneous decoupling limit) $f_{1\nu} = \frac{\rho_{1\nu}}{\rho_{\gamma}} =  \frac{7}{8}\left(\frac{4}{11}\right)^{4/3} \approx 0.23$.
For $n_b>-3$, we can estimate the redshift of DM decoupling from baryons by equating the previous expressions of $\Gamma_{\DM\text{--}b}$ (in the DM--baryon tight-coupling limit where $T_\DM=T_\gamma$) and $\H$. One finds
\begin{align}
1+z_{\DM\text{--}b}^{n_b>-3} = 40\,000
& \left[
\frac{9.15}{c_{n_b}}
\left(
\frac{1+ N_\mathrm{eff}f_{1\nu}}{1+3.044\, f_{1\nu}}
\right)^{1/2} \!\!\!
\left(\frac{\omega_b}{0.0224}\right)^{-1} \!\!\!
\frac{R_\DM^\frac{n_b+1}{2}}{(1+R_\DM)^\frac{n_b-1}{2}} \right. \nonumber \\
&\left.
\left(\frac{\mathcal{F}_\mathrm{He}}{0.76}\right)^{-1}
\left(\frac{\sigma_{\DM\text{--}b}}{10^{4n_b-25} \mathrm{cm}^2}\right)^{-1}
\right]^\frac{2}{n_b+3}\,.
\label{eq:z_dec_dm_b}
\end{align}
In the result section (Sec.~\ref{sec:res}), we see that CMB bounds are of the order of magnitude of ${\sigma_{\DM\text{--}b}} \sim {\cal O}({10^{4n_b-25} \mathrm{cm}^2})$. Thus, the term between brackets is always of order one or bigger. This means that for $n_b>-3$, DM decouples from baryons around $z_{\DM\text{--}b} \sim {\cal O}(40\,000)$ or earlier, hence during radiation domination. 
Equations (\ref{eq:zdecb2}, \ref{eq:zdecb0}) are the restrictions of (\ref{eq:z_dec_dm_b}) to the case $n_b=-2$ and $n_b=0$.
For $n_b=-4$, we can compare the rate $\Gamma_{\DM\text{--}b}$ with $\H$ at $z\sim 10^4$, when the parenthesis in equation~(\ref{eq:coldmb}) is still dominated by $T_b/m_b\sim T_\gamma/m_b$.
This gives 
\begin{equation}
\frac{\Gamma_{\DM\text{--}b}}{\H} \sim 0.06
\left(\frac{\omega_b}{0.0224}\right) 
\left(
\frac{1+ N_\mathrm{eff}f_{1\nu}}{1+3.044\, f_{1\nu}}
\right)^{-1/2} 
\left(1+R_\DM \right)^{\frac{5}{2}} 
\left(\frac{\mathcal{F}_\mathrm{He}}{0.76}\right)
\left(\frac{\sigma_{\DM\text{--}b}}{10^{-41} \mathrm{cm}^2}\right) \, ,
\label{eq:decb4}
\end{equation}
which shows that for typical allowed models, DM and baryons recouple at the earliest around the time of photon decoupling, when $z\sim{\cal O}(10^3)$.

\subsection*{Dark matter -- photon decoupling\label{sec:decb}}

The conformal DM--photon momentum exchange rate is given by equation~(\ref{eq:gamma_dm_g}) and the DM--photon cross section can be parametrised with equation~(\ref{eq:gamma_dm_g2}). Evaluating the various factors gives
\begin{equation}
\Gamma_{\DM\text{--}\gamma} = 6.97\times10^{-30}  (1+z)^3 \left(\frac{u_{\DM\text{--}\gamma}}{10^{-4}}\right) a_0 \, \mathrm{s}^{-1}\, .
\end{equation}
Given the expression (\ref{eq:aHrd}) for the conformal Hubble rate during radiation domination, the DM decouples from photons when the redshift is
\begin{equation}
1+z_{\DM\text{--}\gamma} = 5.48\times10^{4}  \left(
\frac{1+ N_\mathrm{eff}f_{1\nu}}{1+3.044\, f_{1\nu}}
\right)^{1/4} \left(\frac{u_{\DM\text{--}\gamma}}{10^{-4}}\right)^{-1/2} \, .
\end{equation}
In the result section, we see that CMB bounds are of the order of magnitude of \mbox{$u_{\DM\text{--}\gamma} \sim {\cal O}(10^{-4})$}, implying that this decoupling takes place during radiation domination, when \mbox{$z\simeq{\cal O}(10^4)$} or earlier. The conformal photon--DM momentum exchange rate reads
\begin{equation}
\Gamma_{\gamma\text{--}\DM} = \frac{3 \rho_\DM}{4 \rho_\gamma} \Gamma_{\DM\text{--}\gamma} = 2.54\times10^{-26} \left(\frac{\omega_\DM}{0.12}\right) (1+z)^2  \left(\frac{u_{\DM\text{--}\gamma}}{10^{-4}}\right) a_0 \, \mathrm{s}^{-1}\, ,
\end{equation}
such that photons decouple from DM even earlier, when
\begin{equation}
1+z_{\gamma\text{--}\DM} = 8.27\times10^5  \left(
\frac{1+ N_\mathrm{eff}f_{1\nu}}{1+3.044\, f_{1\nu}}
\right) \left(\frac{\omega_\mathrm{M}}{0.12}\right)^{-1}  \left(\frac{u_{\DM\text{--}\gamma}}{10^{-4}}\right)^{-1}\,.
\label{eq:decgdm}
\end{equation}

\subsection*{Dark matter -- dark radiation decoupling}

For $n_\DR > 0$, by equating the rate in equation~(\ref{eq:gammadmdrpar}) with the conformal Hubble rate $\H$ during radiation domination (equation~(\ref{eq:aHrd})), one gets an approximation for the redshift at which DM decouples from~DR:
\begin{equation}
1+z_{\DM\text{--}\DR} = 10^{8+\frac{2}{n_\DR}} \left[ 
\left(
\frac{1+ N_\mathrm{eff}f_{1\nu}}{1+3.044\, f_{1\nu}}
\right)^\frac{1}{2}
\left(
\frac{\Gamma_{\DM\text{--}\DR}^0}{10^{-8n_\DR-22} \, \mathrm{s}^{-1}}
\right)^{-1}
\right]^\frac{1}{n_\DR}\,,
\label{eq:dmdrdec_app}
\end{equation}
where $N_\mathrm{eff}$ is usually given by $3.044+\Delta N_\DR$ and includes the additional DR contribution.
For typical values of $\Gamma_{\DM\text{--}\DR}^0$ compatible with observations,
$\Gamma_{\DM\text{--}\DR}^0 < 10^{-8n_\DR-22} \, \mathrm{s}^{-1}$~\cite{Archidiacono:2019wdp},
 the term between brackets is larger than one, and equation (\ref{eq:dmdrdec}) shows that decoupling takes place during radiation domination. The same is true for the decoupling of DR from DM, which depends on
\begin{equation}
\Gamma_{\DR\text{--}\DM} = \frac{3 \rho_\DM}{4 \rho_\DR} \Gamma_{\DM\text{--}\DR} = 1.60 \cdot 10^4 \left(\frac{\omega_\DM}{0.12}\right) (\Delta N_\DR)^{-1} 
\Gamma_{\DM\text{--}\DR}^0 a_0 \left({1+z}\right)^{n_\DR}\,.
\end{equation}
That decoupling occurs around
\begin{equation}
1+z_{\DR\text{--}\DM} = 10^{8-\frac{2}{1+n_\DR}} \left[ \frac{1.60}{\Delta N_\DR}
 \left(\frac{\omega_\DM}{0.12}\right)
\left(
\frac{1+ N_\mathrm{eff}f_{1\nu}}{1+3.044\, f_{1\nu}}
\right)^\frac{1}{2}
\left(
\frac{\Gamma_{\DM\text{--}\DR}^0}{10^{-8n_\DR-22} \, \mathrm{s}^{-1}}
\right)^{-1}
\right]^\frac{1}{1+n_\DR} \,.
\label{eq:dmdrdec_app2}
\end{equation}
Finally, for $n_\DR =0$, we can express the ratio $\Gamma_{\DM\text{--}\DR}/\H$ during radiation domination as
\begin{equation}
\frac{\Gamma_{\DM\text{--}\DR}}{\H} = 4.76 \cdot 10^{-2} 
\left(
\frac{1+ N_\mathrm{eff}f_{1\nu}}{1+3.044\, f_{1\nu}}
\right)^{-\frac{1}{2}}
\left(
\frac{\Gamma_{\DM\text{--}\DR}^0}{10^{-21} \, \mathrm{s}^{-1}}
\right)\, .
\label{eq:dmdrdec_app3}
\end{equation}
In this case, CMB bounds are of the order of  ${\Gamma_{\DM\text{--}\DR}^0} \leq {\cal O}(10^{-21}) \, \mathrm{s}^{-1} \sim {\cal O}(10^{-7}) \, \mathrm{Mpc}^{-1}$, showing that $\epsilon$ is at most of the order of $10^{-1}$ during radiation domination.

%% file: App_Implementation.tex
\section{Implementation in \class}\label{App:num}

As usual in \class, each new species is identified by a short acronym, which allows for a quick search of all of the relevant equations (described in App.~\ref{App:eqs}). As we are considering only one IDM species with multiple interactions (as discussed in Sec.~\ref{sec:theory}), the relevant species are:\begin{itemize}
\item \texttt{idm} $\longrightarrow$ interacting dark matter species 
\item \texttt{idr} $\longrightarrow$ interacting dark radiation 
\end{itemize}
Additionally, to find the specifics of each type of interaction for the IDM, the following acronyms are employed:\begin{itemize}
\item \texttt{idm\_b} $\longrightarrow$ interacting dark matter--baryon
\item \texttt{idm\_g} $\longrightarrow$ interacting dark matter--photon
\item \texttt{idm\_dr} $\longrightarrow$ interacting dark matter--dark radiation
\end{itemize}

In order to follow the full temperature evolution of the IDM species, its temperature needs to be integrated together with the baryon temperature, (as discussed in Sec.~\ref{sec:th_idm}) which was not done in \class until now, as it was not necessary. As this is a stiff system of equations, it requires an \texttt{ndf15} integrator, which was already present in the \texttt{perturbations} module, and was incorporated in the \texttt{background} and \texttt{thermodynamics} modules in \class v3.0~\cite{Lucca:2019rxf}. Additionally, as described in Sec.~\ref{sec:th_idm}, \class now has several criteria to choose the appropriate initial conditions for the DM temperature evolution.

\subsection*{Input parameters  \texttt{idm}}
\begin{table}
	\centering
	\hspace*{-3em}
	\begin{tabular}{| l | c | c | c |}
		\hline
		This work & \textsc{class v3.1} & \textsc{class v2.9}  \\ \hline
		$m_\DM$ & \texttt{m\_idm} & -- \\ \hline
		$\sigma_{\DM\text{--}b}$ & \texttt{cross\_idm\_b} & -- \\
		$n_b$ & \texttt{n\_index\_idm\_b} & -- \\ \hline
		$\sigma_{\DM\text{--}\gamma}$ & \texttt{cross\_idm\_g} & -- \\
		$u_{\DM\text{--}\gamma}$& \texttt{u\_idm\_g} & -- \\ \hline
		$a_\mathrm{dark}$ & \texttt{a\_idm\_dr} & \texttt{a\_dark}\\
		$\Gamma^0_{\DM\text{--}\DR}$ & \texttt{Gamma\_0\_idm\_dr}  & \texttt{Gamma\_0\_nadm}\\
		$N_\DR$ & \texttt{N\_idr} & \texttt{N\_dg}\\
		$\xi$ & \texttt{xi\_idr} & \texttt{xi\_idr}\\
		$n_\DR$ & \texttt{n\_index\_idm\_dr} &  \texttt{nindex\_dark} \\
 \hline 
	\end{tabular}
	\caption{Correspondence between the notation of this work and the input parameters for \textsc{class v3.1} and \textsc{class v2.9} (for DM--DR interactions). 
	}
	\label{tab:notations2}
\end{table}
We summarise the correspondence between the input parameters used in \class and the notation used in this work in table~\ref{tab:notations2}.

\noindent The following parameters control the overall properties of the IDM species: \begin{itemize}
\item \texttt{f\_idm}: fraction of DM that will be interacting (default 0). Can also be passed in the form \texttt{Omega\_idm} or \texttt{omega\_idm}.
\item \texttt{m\_idm}: mass of the interacting DM particle, in eV (default $10^9$)
\end{itemize}
\subsection*{Input parameters \texttt{idm\_b}}
For the specific DM--baryon interactions, the code requires the following two quantities: \begin{itemize}
\item \texttt{cross\_idm\_b}: coupling strength between the DM and baryons in cm$^2$ (default 0)
\item \texttt{n\_index\_idm\_b}: temperature dependence of the DM--baryon interactions, between $-4$ and $4$ (default 0)
\end{itemize}

\subsection*{Input parameters  \texttt{idm\_g}}
For the specific DM--photon interactions, the code requires only the following quantity: \begin{itemize}
\item \texttt{cross\_idm\_g}: coupling strength between the DM and baryons in cm$^2$ (default 0). Can also be passed as the relative cross section (see eq.~\ref{eq:dmg}) in the form \texttt{u\_idm\_g}
\end{itemize}

\subsection*{Input parameters  \texttt{idm\_dr} and  \texttt{idr}}
The input parameters related to DM--DR interactions are described in detail in the corresponding release paper of \class v2.9~\cite{Archidiacono:2019wdp}, in Sec. 3.2 therein.

However, to homogenise the notation of the different IDM interactions, we have renamed several of these parameters: this is shown in table~\ref{tab:notations2}. Nonetheless, \class also accepts as input the old notation from Refs.~\cite{Archidiacono:2017slj,Archidiacono:2019wdp}.

\begin{table}
	\centering
	\hspace*{-0em}
	\begin{tabular}{| l | c | c |}
		\hline
		Case & Runtime  [s] & \% Slowdown \\ \hline
		\lcdm	&	0.936	&	0.0	\\ \hline
		DM--b $(n_b = -4)$	&	0.959	&	2.4	\\
		DM--b $(n_b = -2)$	&	0.949	&	1.4	\\
		DM--b $(n_b = 0)$	&	0.950	&	1.5	\\ \hline
		DM--$\gamma$	&	0.940	&	0.4	\\ \hline
		DM--DR, $n_\DR =0$, fluid DR 	         & 1.307		& 39.6  \\ 
		DM--DR, $n_\DR =0$, free-streaming DR & 2.181 	& 132.9 \\ 
		DM--DR, $n_\DR =4$, fluid DR          & 1.622 	& 73.3	\\ 
		DM--DR, $n_\DR =4$, free-streaming DR & 4.082 	& 336.0\\ \hline
		DM--b $(n_b = -4)$+DM--$\gamma$	&	0.994	&	6.1	\\
		DM--b $(n_b = -2)$+DM--$\gamma$	&	0.982	&	4.9	\\
		DM--b $(n_b = 0)$+DM--$\gamma$	&	0.983	&	4.9	\\ \hline
		DM--b $(n_b = -4)$+DM--DR, $n_\DR =0$, fluid DR	&	1.360	&	45.3	\\
		DM--b $(n_b = -2)$+DM--DR, $n_\DR =0$, fluid DR	&	1.374	&	46.7	\\
		DM--b $(n_b = 0)$+DM--DR, $n_\DR =0$, fluid DR	&	1.340	&	43.1	\\ \hline
		DM--$\gamma$+DM--DR	&	1.356	&	44.8	\\ \hline
		DM--b $(n_b = -4)$+DM--$\gamma$+DM--DR, $n_\DR =0$, fluid DR	&	1.590	&	69.8	\\
		DM--b $(n_b = -2)$+DM--$\gamma$+DM--DR, $n_\DR =0$, fluid DR	&	1.378	&	47.2	\\
		DM--b $(n_b = 0)$+DM--$\gamma$+DM--DR, $n_\DR =0$, fluid DR	&	1.396	&	49.1	\\
 \hline 
	\end{tabular}
	\caption{Average performance for all of the DM interaction models considered here.}
	\label{tab:speed}
\end{table}

\subsection*{Code Performance}
In table~\ref{tab:speed} we show the average runtime of the code for the different interacting models, using the $2\sigma$ limits from table~\ref{tab:bounds} for the interaction rates, $m_\DM=1$\,GeV and $N_\DR=0.5$ (if DM--DR interactions are active). All runtime checks were performed on 8 CPUs on a Dell XPS with Intel Core i7-8665U CPU (1.90GHz).

It is worth pointing out that none of the interacting models considered in this work cause a significant slowdown of the code, at most slowing it down by $\sim 70\%$, and remaining always under $1.5\mathrm{s}$ runtime. 

We have additionally included the runtime on different DR models that are not considered in this work, but that can be treated by the code. When assuming that the DR is free-streaming instead of behaving like a fluid, the full Boltzmann hierarchy needs to be considered, which slows down the code by a factor $\sim 1.5$ compared to the case with fluid DR (for $n_\DR=0$). There is a further slowdown when going from $n_\DR=0$ to $n_\DR =4$ (likewise for $n_\DR =2$), due to the DM and DR species being tightly coupled in the early universe. However, even in the slowest scenario of free-streaming DR with $n_\DR =4$, the code is only a factor $\sim 4$ slower than for the \lcdm~model.